\documentclass[12pt,a4paper]{iopart}

\usepackage[english]{babel}

\usepackage{iopams}
\usepackage{verbatim}

\usepackage{graphicx}

\usepackage{color}

\usepackage[colorlinks,citecolor=blue,linkcolor=black,urlcolor=blue]{hyperref}

\def\iotabar{\lower3pt\hbox{$\mathchar'26$}\mkern-7mu\iota}
\newcommand {\aplt} {\ {\raise-.5ex\hbox{$\buildrel<\over\sim$}}\ }
\newcommand{\dd}{\mbox{d}}

\newcommand{\eq}[1]{(\ref{#1})}
\newcommand{\bun}{\hat{\mathbf{b}}}
\newcommand{\eun}{\hat{\mathbf{e}}}

\newcommand{\boldr}{\mathbf{r}}

\newcommand{\bv}{\mathbf{v}}

\newcommand{\bR}{\mathbf{R}}

\newcommand{\bB}{\mathbf{B}}

\newcommand{\dotcross}{ \raise 0.65ex\hbox{${\scriptstyle {{_{\displaystyle \cdot}}\atop\times}}$} }
\newcommand{\crossdot}{ \raise 0.5ex\hbox{${\scriptstyle {{_\times}\atop{\displaystyle \cdot}}}$} }

\newcommand{\kappabf}{\mbox{\boldmath$\kappa$}}

\newcommand{\cE}{{\cal E}}

\newcommand{\sumsig}{ \raise -1.3ex\hbox{${{\displaystyle \sum}\atop{\scriptstyle \sigma}}$} }
\newcounter{appnumb}

\newcommand{\gbl}{{g_{\rm bl}}}

\begin{document}

\title[The effect of tangential drifts in stellarators close to
omnigeneity] {The effect of tangential drifts on neoclassical transport in stellarators 
close to
  omnigeneity}

\author{Iv\'an Calvo$^{1}$}
\vspace{-0.2cm}
\eads{\mailto{ivan.calvo@ciemat.es}}
\vspace{-0.6cm}
\author{Felix I Parra$^{2,3}$}
\vspace{-0.2cm}
\eads{\mailto{felix.parradiaz@physics.ox.ac.uk}}
\vspace{-0.6cm}
\author{Jos\'e Luis Velasco$^{1}$}
\vspace{-0.2cm}
\eads{\mailto{joseluis.velasco@ciemat.es}}
\vspace{-0.6cm}
\author{J Arturo Alonso$^{1}$}
\vspace{-0.2cm}
\eads{\mailto{arturo.alonso@ciemat.es}}

\vspace{0.5cm}

\address{$^1$Laboratorio Nacional de Fusi\'on, CIEMAT, 28040 Madrid, Spain}
\address{$^2$Rudolf Peierls Centre for Theoretical Physics, University of Oxford, Oxford, OX1 3NP, 
UK}
\address{$^3$Culham Centre for Fusion Energy, Abingdon, OX14 3DB, UK}



\vspace{-0.2cm}

\begin{abstract}
  In general, the orbit-averaged radial magnetic drift of trapped
  particles in stellarators is non-zero due to the three-dimensional
  nature of the magnetic field. Stellarators in which the
  orbit-averaged radial magnetic drift vanishes are called
  omnigeneous, and they exhibit neoclassical transport levels
  comparable to those of axisymmetric tokamaks. However, the effect of deviations from omnigeneity cannot be neglected in
  practice, and it is more deleterious at small
  collisionalities. For sufficiently low collision frequencies (below the values that define the 
$1/\nu$ regime), the components of the drifts 
tangential to the flux surface become relevant. This article focuses on the study of such 
collisionality regimes in stellarators close to omnigeneity when the gradient of the non-omnigeneous perturbation is small. First, it is proven that closeness to omnigeneity is required to actually preserve radial locality in the drift-kinetic equation for collisionalities below the $1/\nu$ regime. Then, using the derived radially local equation, it is shown that neoclassical transport is determined by two layers located at different regions of phase space. One
  of the layers corresponds to the so-called $\sqrt{\nu}$ regime and
  the other to the so-called superbanana-plateau regime. The importance of the superbanana-plateau layer for the calculation of the tangential electric field is emphasized, as well as the relevance of the latter for neoclassical transport in the collisionality regimes considered in this paper. In particular, the role of the tangential electric field is essential for the emergence of a new subregime of superbanana-plateau transport when the radial electric field is small. A formula for the
  ion energy flux that includes the $\sqrt{\nu}$ regime and
  the superbanana-plateau regime is given. The energy flux scales with the square of the size of the deviation from omnigeneity. Finally, it is explained why below a certain collisionality value the formulation presented in this article ceases to be 
valid.
\end{abstract}

\maketitle

\section{Introduction}
\label{sec:introduction}

Stellarators~\cite{Helander2012} offer some intrinsic advantages
compared to tokamaks, such as the possibility of steady-state
operation and the absence of disruptions. However, the magnetic
configuration of a stellarator has to be designed very carefully for
it to have confinement properties comparable to those of an
axisymmetric tokamak. In a generic stellarator, trapped particle
orbits have non-zero secular radial drifts and they leave the device
in a short time. The stellarator configuration is called
 omnigeneous~\cite{Cary1997a,Cary1997b,Parra2015,Landreman2012} if the secular radial drifts
vanish.

Omnigeneity guarantees a neoclassical transport level similar to that in a tokamak (see equation \eq{eq:sizeQiomnigeneous} below). Define the normalized ion
gyroradius $\rho_{i*} := v_{ti}/(\Omega_iL_0)$, where $v_{ti}$ and
$\Omega_i$ are the ion thermal speed and the ion gyrofrequency, and
$L_0$ is the typical length of variation of the magnetic field, which is assumed to be
of the order of the system size. The
gyrofrequency is $\Omega_i = Z_ie B/(m_i c)$, where $Z_i e$ is the
charge of the ions, $e$ is the elementary charge, $B$ is the magnitude
of the magnetic field $\bB$, $m_i$ is the ion mass, and $c$ is the speed of
light. Since $\rho_{i*} \ll 1$ in a strongly magnetized plasma, the
drift-kinetic formalism~\cite{Hazeltine73} is appropriate. Denoting by $f_i(\boldr,\bv)$ the 
phase-space distribution, the radial ion energy flux $Q_i$ reads
\begin{equation}\label{eq:defenergyfluxIntro}
Q_i = \int \dd^2 S\int \dd^3v
\left(\frac{m_iv^2}{2}
+ Z_i e \varphi
\right)
\bv_{d,i}\cdot\hat\mathbf{n}\,
 f_{i}.
\end{equation}
Here, $\varphi$ is the electrostatic potential, $\bv_{d,i}$ is the drift velocity, $\hat\mathbf{n}$ is the unit
vector normal to the flux surface, $\bv_{d,i}\cdot\hat\mathbf{n} \sim
\rho_{i*} v_{ti}$, and the integrals are performed over velocity space
and over the flux surface. In a perfectly omnigeneous
stellarator $f_i$, to lowest order in $\rho_{i*}$, is a Maxwellian $f_{Mi}$ with density $n_i$ and
temperature $T_i$ that are constant on flux surfaces. The phase-space distribution is written as $f_i = f_{Mi} + f_{i1}$, where the perturbation to the Maxwellian is found to have a size $f_{i1} \sim O(\rho_{i*} f_{Mi})$. The first non-vanishing contribution to the energy flux comes from a piece of the distribution function that is $O(\nu_{i*}\rho_{i*} f_{Mi})$, where $\nu_{i*} := \nu_{ii}L_0/v_{ti}$ is the ion collisionality and
$\nu_{ii}$ is the ion-ion collision frequency. Then, in an omnigeneous stellarator,
\begin{equation}\label{eq:sizeQiomnigeneous}
Q_i \sim \nu_{i*}\rho_{i*}^2 n_iT_iv_{ti}S_\psi.
\end{equation}
The area of the flux
surface is denoted by $S_\psi$, with $\psi$ the radial coordinate.

The proof of Cary and Shasharina~\cite{Cary1997a,Cary1997b} for the
existence of omnigeneous magnetic fields implies that exact omnigeneity throughout the plasma requires, at least, non-analiticity. Let us explain
this in more detail. As shown in references \cite{Cary1997a} and
\cite{Cary1997b}, there exist omnigeneous magnetic fields that are
analytic. These configurations coincide with the set of quasisymmetric
magnetic fields~\cite{Boozer83, Nuehrenberg88}. To the virtues of
omnigeneity, quasisymmetry adds the vanishing of neoclassical flow damping
in the quasisymmetric direction. Therefore, in quasisymmetric
stellarators larger flow velocities can be attained. In principle,
a quasisymmetric stellarator plasma may have large flow
shear, that in principle can reduce turbulent
transport~\cite{Connor2004}. However, the quasisymmetry condition is
incompatible with the magnetohydrodynamic equilibrium equations~\cite{Garren1991}, and the stellarator can be made
quasisymmetric only in a limited radial region.

This is why we said above that a necessary condition
for exact omnigeneity is non-analiticity; specifically, the
discontinuity of some derivatives of second or higher order. However,
designing and aligning coils that create a magnetic field with
discontinuous derivatives at certain points in space is probably
technically impossible. Therefore, even in optimized magnetic fields, the effect of deviations
from the desired omnigeneous configuration cannot be neglected. It is
thus necessary to study magnetic fields of the form $\bB = \bB_0
+ \delta \bB_1$, where $\bB_0$ is omnigeneous and $\delta \bB_1$
is a perturbation, with $0\leq \delta \ll 1$ and $\bB_1 \sim
\bB_0$.

The effect of a deviation from omnigeneity is more detrimental for confinement at small
collisionalities.  If $\nu_{i*} \ll 1$ and the stellarator is non-omnigeneous,
the non-omnigeneous piece of $f_{i1}$ becomes large, so that $f_{i1}
\gg \rho_{i*} f_{Mi}$ and the energy flux can be much larger than the
estimation \eq{eq:sizeQiomnigeneous} even if $\delta$ is small. The
quantification of this effect for
\begin{equation}\label{eq:def1overnuregime}
\rho_{i*}\ll\nu_{i*}\ll 1,
\end{equation}
that defines the $1/\nu$ regime, has been treated in \cite{Calvo13, Calvo14, Calvo15} for 
stellarators close to quasisymmetry and is the subject of \cite{Parra2014} 
for stellarators close to omnigeneity.
However, this regime does not exhaust the low collisionality parameter
space in stellarators. When
\begin{equation}\label{eq:defVeryLowCollisionality}
\nu_{i*} \lesssim \rho_{i*},
\end{equation}
the components of the drifts tangential to the 
flux surface matter~\cite{Helander2012, Beidler2011, Logan2013, Matsuoka2015}. In
 this paper we study stellarators close to omnigeneity in the collisionality
regime \eq{eq:defVeryLowCollisionality}, relevant
for a stellarator reactor~\cite{Dinklage2013}.

It is important to point out that the calculations in this paper do
not rely on large aspect ratio approximations. Of course, if the
stellarator close to omnigeneity under consideration has large aspect
ratio, one can perform a subsidiary expansion in the inverse aspect
ratio and refine the results obtained here. This will be the subject of future
work. The rest of the paper is organized as follows.

In Section \ref{sec:OmnigeneousStellarators} we introduce a set of
flux coordinates that is well-adapted to stellarator magnetic
geometries. Then, we give the formal definition of omnigeneity.

In Section \ref{sec:ExpansionDKEclosetoOmnigeneity} we derive,
starting from the complete drift-kinetic equation, the equation for
the dominant component of the distribution function when $\delta\ll
1$ and $\nu_{i*} \lesssim \rho_{i*}$. In particular, we explain why the
standard expansion in $\rho_{i*}$ breaks down for a
generic stellarator when $\nu_{i*} \lesssim \rho_{i*}$. In brief, the reason
is that $f_{i1}$ becomes so large that $f_{i1}\sim f_{Mi}$. For
stellarators close to omnigeneity, however, we can expand in 
the small parameter $\delta$. In addition, in a generic stellarator the drift-kinetic 
equation becomes radially non-local when $\nu_{i*} \lesssim \rho_{i*}$, but we will see that with the 
condition $\delta\ll 1$ we can derive a radially local drift-kinetic equation in this collisionality 
regime.

A precision must be made about the asymptotic expansion in $\delta$
carried out in this paper. When $\rho_{i*}\ll\nu_{i*}\ll 1$, it has been
understood (in
\cite{Calvo13, Calvo14, Calvo15} for stellarators close to quasisymmetry and in \cite{Parra2014} for 
stellarators close to omnigeneity) that the
effect of the deviations (from quasisymmetry or omnigeneity) is very
different depending on the size of the gradients on the surface of the
magnetic field perturbation. For the regime
\eq{eq:defVeryLowCollisionality}, the case of deviations with small
gradients and the case of deviations with large gradients also require
different treatments, in principle. Here, we restrict ourselves to deviations with small
gradients. Let us be more
precise. If $B_0 := |\bB_0|$ and $B_1 := |\bB_1|$, by ``deviations with small
gradients" we mean that
${|\nabla \ln B_0|}/{|\nabla \ln B_1|} \gg \delta$. If this inequality is well satisfied, then we can consider that the characteristic lengths of both, $B_0$ and $B_1$, are $O(L_0)$ as far as the asymptotic expansion in $\delta$ is concerned.

In Section \ref{sec:solutionDKEverylowcollisionality} the equation
derived in Section \ref{sec:ExpansionDKEclosetoOmnigeneity} for the
non-omnigeneous piece of the distribution function is solved when $\nu_{i*} \ll
\rho_{i*}$. We find that $Q_i$ is dominated by two
collisional layers in phase space. One of the layers lies at the
boundary between trapped and passing trajectories and produces an
energy flux
\begin{equation}\label{eq:sizeQisqrtnu}
Q_i\sim \delta^2\frac{\nu_{ii}^{1/2}}{\omega_\alpha^{3/2}}
\sqrt{
\ln\left(\omega_\alpha / \nu_{ii}\right)
}
\rho_{i*}^2
n_iT_iv_{ti}^2 L_0^{-1}S_\psi,
\end{equation}
where $\omega_\alpha\sim \rho_{i*} v_{ti} / L_0$, defined in Section  \ref{sec:solutionDKEverylowcollisionality}, is the 
precession frequency due to the tangential drifts. On the right side of \eq{eq:sizeQisqrtnu} $\omega_\alpha$ actually stands for the value of the precession frequency evaluated at the boundary between trapped and passing particles and at $v = v_{ti}$. Note the logarithmic correction to the scaling with $(\nu_{ii}/\omega_\alpha)^{1/2}$ in \eq{eq:sizeQisqrtnu}, that we calculate in subsection \ref{sec:sqrtnuRegime}.

The other layer lies at 
the points of
phase space where $\omega_\alpha$ vanishes and yields $Q_i$ independent of $\nu_{i*}$. 
Namely,
\begin{equation}\label{eq:sizeQisuperbananaplateau}
Q_i\sim \delta^2\rho_{i*} n_iT_iv_{ti}S_\psi.
\end{equation}
The first layer (see subsection \ref{sec:sqrtnuRegime}) gives the so-called $\sqrt{\nu}$ regime,
found in certain models of stellarator geometry~\cite{Galeev1979,Ho1987}
where the inverse aspect ratio and the helical ripple are employed as
expansion parameters. The second layer (see subsection \ref{sec:superbananaplateauRegime}) gives the
superbanana-plateau regime, derived in \cite{Shaing2015} for finite aspect ratio tokamaks 
with broken symmetry. Here, the $\sqrt{\nu}$ and superbanana-plateau regimes are derived and analyzed in a much broader setting and in deeper detail than previously available in the literature.
In particular, we will show that the treatment of the superbanana-plateau regime requires special care for small values of the radial electric field (see \eq{eq:def_smallE_r} for a precise definition of what `small' means in this context), and in those cases logarithmic corrections appear in \eq{eq:sizeQisuperbananaplateau} as well. Of course, neither the $\sqrt{\nu}$ regime nor the superbanana-plateau regime (nor the $1/\nu$ regime) exist in perfectly omnigeneous stellarators; i.e. when $\delta = 0$.

From the start, it will be evident that the role of the electric field tangent to the flux surface is relevant when the collisionality is as low as in \eq{eq:defVeryLowCollisionality} (not to mention its importance for impurity transport, as pointed out, for example, in \cite{Garcia-Regaña2013, Alonso15, Alonso16}). Furthermore, we will show that writing the contributions to the quasineutrality equation that gives the electric field tangent to the flux surface is a subtle issue. In particular, we will prove that the superbanana-plateau layer has to be resolved to find the tangent electric field. For this reason, we discuss the quasineutrality equation in subsection \ref{sec:quasineutrality_equation}.

The contributions to $Q_i$ from the two layers mentioned above are 
additive, as long as the layers are distinct and do not overlap, and a general 
expression embracing the $\sqrt{\nu}$ and superbanana-plateau regimes is provided in subsection \ref{sec:additive formula}. The treatment of cases in which both layers overlap is left for the future. As \eq{eq:sizeQisqrtnu} and \eq{eq:sizeQisuperbananaplateau} already indicate, we will show that the neoclassical fluxes scale with the square of the size of the deviation from omnigeneity, $\delta$.

In Section \ref{sec:varphi1andvarphi0} we use the results of previous sections to write the equation
that gives the radial electric field.

Finally, in Section \ref{sec:Estimationnucritical}, we explain that
the results of Section \ref{sec:solutionDKEverylowcollisionality} are
not expected to be correct for arbitrarily small $\nu_{i*}$. For each $\delta$ there
exists a value of the collisionality $\nu_{\delta *}$ such that if
$\nu_{i*} < \nu_{\delta *}$ our solution is not valid. We explain and
estimate the limit value $\nu_{\delta *}$.

In Section \ref{sec:conclusions} we summarize the conclusions of the
paper.

\section{Omnigeneous stellarators}
\label{sec:OmnigeneousStellarators}

Throughout the paper, we deal with stellarators whose magnetic field
configurations possess nested flux surfaces. In the first place, we
define a set of spatial coordinates $\{\psi,\alpha,l\}$ adapted to the
magnetic field. The coordinate $\psi$
determines the flux surface, whereas $\alpha$ is an angular coordinate that
labels a magnetic field line once $\psi$ has been fixed. Finally $l$,
the arc length over the magnetic field line, specifies the position
along the line for fixed $\psi$ and $\alpha$. Denote by
$\psi(\boldr)$, $\alpha(\boldr)$ and $l(\boldr)$ the functions giving
the value of these coordinates for each point $\boldr$ in the
stellarator. The magnetic field can be written as
\begin{equation}
\bB = \Psi'_t(\psi)\nabla\psi\times\nabla\alpha.
\end{equation}
Here, $\Psi_t$ is the toroidal magnetic flux over $2\pi$ and primes
stand for differentiation with respect to $\psi$. In order to have
unique pairs $(\alpha,l)$ associated to each point on a flux surface,
we choose a curve ${\cal C}$ that closes poloidally\footnote{To fix ideas, we are thinking of $
\alpha$ as a poloidal angle, but things work analogously if $\alpha$, and therefore the curve ${\cal C}$, 
have a different helicity.}. This 
curve can be
parameterized by $\alpha$. All points on the curve are assigned, by
definition, the value $l=0$. For each pair $\psi$ and $\alpha$ we
take $l\in[0,L(\psi,\alpha))$, where $L(\psi,\alpha)$ is found by
integrating from ${\cal C}$ along the line until the curve ${\cal C}$
is encountered again.

Let $v$ be the magnitude of the velocity and $\lambda =
v_\perp^2/(v^2B)$ the pitch angle. Given a flux surface determined by
$\psi$, particles are passing or trapped depending on the value of
$\lambda$. Passing trajectories have $\lambda < 1/B_{\rm max}(\psi)$,
where $B_{\rm max}(\psi)$ is the maximum value of $B$ on the flux
surface. Passing particles explore the entire flux surface and always
have vanishing average radial magnetic drift. Particles with $\lambda
> 1/B_{\rm max}(\psi)$ are trapped. For trapped particles, the radial
magnetic drift averaged over the orbit is non-zero in a generic
stellarator. A stellarator is called omnigeneous if the orbit-averaged
radial magnetic drift is zero for all
particles~\cite{Cary1997a,Cary1997b,Parra2015,Landreman2012}. That is, if and only
if the second adiabatic invariant $J = 2\int_{l_{b_1}}^{l_{b_2}} |v_{||}| \dd l$ is a flux function, 
which means that
\begin{equation}\label{eq:definitionomnig}
\partial_\alpha \int_{l_{b_1}}^{l_{b_2}}\sqrt{1-\lambda B}\dd l = 0
\end{equation}
must hold for every trapped trajectory. Here $l_{b_1}$ and $l_{b_2}$ are the bounce 
points; i.e. the
solutions for $l$ of the equation $1-\lambda B(\psi,\alpha,l) = 0$ for a
particular trapped trajectory. Since \eq{eq:definitionomnig} has to be
satisfied for every $\lambda$, that equation is equivalent to requiring\footnote{In \cite{Cary1997b}, it is proven that  \eq{eq:definitionomnig} implies that $\sum |\dd l / \dd B|$ (the sum runs over the two points of each well where the magnitude of the magnetic field reaches a certain value $B$) depends only on $\psi$ and on the value of $B$. Property \eq{eq:definitionomnigEquiv} follows by employing this result after changing the integration variable on the left-side of \eq{eq:definitionomnigEquiv} from $l$ to $B$.}
\begin{equation}\label{eq:definitionomnigEquiv}
  \partial_\alpha \int_{l_{b_1}}^{l_{b_2}}\Lambda(\psi,v,\lambda,B(\psi,\alpha,l))
\dd l = 0
\end{equation}
for any function $\Lambda$ that depends on $\alpha$ and $l$ only
through $B$. We will make use of this definition of omnigeneity
several times along the article.

\section{Low-collisionality drift-kinetic equation in 
  stellarators close to omnigeneity}
\label{sec:ExpansionDKEclosetoOmnigeneity}

As we said in the Introduction, due to the smallness of $\rho_{i*}$ we can employ the 
drift-kinetic approach~\cite{Hazeltine73, Calvo13, parra08, ParraCalvo11}. It consists 
of a systematic way to average, order by order
in $\rho_{i*}$, over the fast gyration of particles around magnetic
field lines. This is achieved by finding a coordinate transformation
on phase space that decouples the gyromotion from the comparatively
slow motion of the guiding center. The new coordinates are called
drift-kinetic coordinates. In what follows, we restrict ourselves to electrostatic 
drift-kinetics and assume that $\varphi\sim m_i v_{ti}^2 / e$.

The form of the drift-kinetic equation is determined by the
transformation from coordinates $\{\boldr,\bv\}$ to drift-kinetic
coordinates (or, perhaps more precisely, to the drift-kinetic limit of gyrokinetic coordinates~\cite{Catto78}). Even though we will end up employing the
  coordinates $v$ and $\lambda$ defined in Section
  \ref{sec:OmnigeneousStellarators}, it is convenient to
  start using as independent coordinates the total energy per mass
  unit $\cE$ and the magnetic moment $\mu$ because they are constants of the particle motion. Then, in drift-kinetic coordinates 
  $\{\bR,\cE,\mu,\sigma,
\gamma\}$, where $\bR$ is the position of the guiding center, $\sigma$
is the sign of the parallel velocity and $\gamma$ is the
gyrophase, we have
\begin{eqnarray}
 \bR = \boldr - \frac{1}{\Omega_i}\bun\times \bv 
+ O(\rho_i^{*2}L_0),
\nonumber\\[5pt]
\cE = \frac{v^2}{2} + \frac{Z_ie\varphi}{m_i},
\nonumber\\[5pt]
\mu = \frac{1}{2B}(v^2 - (\bv\cdot\bun)^2) + O(\rho_{i*}v_{ti}^2/B),
\nonumber\\[5pt]
\gamma = \arctan
\left(
\frac{\bv\cdot\eun_2}{\bv\cdot\eun_1}
\right) + O(\rho_{i*}),
\end{eqnarray}
where $\bun = B^{-1}\bB$ and the right sides of the previous expressions are evaluated at
$\boldr$.  The orthogonal unit vector fields
$\eun_1$ and $\eun_2$ satisfy at each point $\eun_1\times\eun_2 =
\bun$. The higher-order corrections in the definition of $\mu$ are
determined by the fact that $\mu$ is the adiabatic invariant
corresponding to the ignorable coordinate $\gamma$. Finally, $\sigma =
v_{||}/|v_{||}|$ gives the sign of the parallel velocity, where the
latter is viewed as a function of the other coordinates through the
expression
\begin{equation}
v_{||} = \sigma \sqrt{2\left(
\cE-\mu B-\frac{Z_ie \varphi}{m_i}
\right)}\ \,.
\end{equation}

Denote by $F_i(\psi(\bR),\alpha(\bR),l(\bR),\cE,\mu,\sigma)$ the
distribution function in drift-kinetic coordinates. We assume from the
beginning that our distribution function does not depend on the
gyrophase $\gamma$, which is true for all the calculations in this paper (see
\cite{parra08} for the proof that only pieces of the distribution function $O(\nu_{i*}\rho_{i*} f_{Mi})$ or smaller are gyrophase dependent). In these
coordinates the drift-kinetic equation reads
\begin{equation}\label{eq:DKEwithEtotal}
\dot\bR \cdot \nabla F_i = C^\cE_{ii}[F_i,F_i].
\end{equation}
Here,
\begin{eqnarray}\label{eq:dotRpar}
\dot\bR\cdot\bun\bun = v_{||}\bun + O(\rho_{i*}v_{ti})
\end{eqnarray}
and
\begin{eqnarray}\label{eq:dotRperp}
  \dot\bR - \dot\bR\cdot\bun\bun =  \bv_{M,i} + \bv_{E}
+ O(\rho_i^{*2}v_{ti}),
\end{eqnarray}
with
\begin{equation}\label{eq:magneticdrift}
\fl\bv_{M,i} = \frac{1}{\Omega_i}\bun\times
\left(
v_{||}^2\kappabf + \mu\nabla B
\right)
\end{equation}
being the magnetic drift,
\begin{equation}\label{eq:ExBdrift0}
\fl\bv_{E} = \frac{c}{B}\bun\times
\nabla\varphi
\end{equation}
being the $E\times B$ drift and $\kappabf = \bun\cdot\nabla\bun$ being the curvature of the magnetic 
field lines. Note that $|\bv_{M,i}|$ and $|\bv_{E}|$ are $O(\rho_{i*} v_{ti})$. 

In \eq{eq:dotRpar} and \eq{eq:dotRperp} we have shown only the terms
that will be needed later on. All the terms of $\dot\bR$ up to
$O(\rho_i^{*2}v_{ti})$ have been computed in \cite{ParraCalvo11}. In
\eq{eq:DKEwithEtotal}, an expansion in the mass ratio
$\sqrt{m_e/m_i}\ll 1$ has been taken so that ion-electron collisions
are neglected, and $C^\cE_{ii}$ is the ion-ion Landau collision
operator written in coordinates $\cE$ and $\mu$. Its explicit
expression (see \cite{helander02bk}, for example) is not necessary for
our purposes. From here on, we concentrate on ion transport.

Low collisionality regimes are defined by $\nu_{i*}\ll 1$.  It is well-known (see, for example, subsection 7.1 in \cite{Calvo13} and also \cite{Parra2014}) that if the collisionality is small but still larger than the normalized gyroradius, i.e. if $\rho_{i*}\ll \nu_{i*}\ll 1$, then the distribution function and electrostatic potential can be expanded as
\begin{equation}\label{eq:expansionF1overnu}
  F_{i}
  = F_{i0}^\cE + F_{i1} + \dots
\end{equation}
and
\begin{equation}\label{eq:expansionvarphi1overnu}
\varphi = \varphi_0 + \varphi_1 + \dots,
\end{equation}
where
\begin{equation}\label{eq:MaxwellianF0Etotal}
F_{i0}^\cE(\psi,\cE) = 
n_i(\psi)\left(\frac{m_i}{2\pi T_i(\psi)}\right)^{3/2}
\exp\left(-\frac{m_i\cE-Z_ie\varphi_0(\psi)}{T_i(\psi)}\right)
\end{equation}
is a Maxwellian distribution with density and temperature constant on the flux surface, the non-adiabatic perturbation to $F_{i0}^\cE$ has a size
\begin{equation}\label{eq:sizeF11overnu}
F_{i1} \sim \frac{\rho_{i*}}{\nu_{i*}} F_{i0}^\cE,
\end{equation}
$\varphi_0(\psi) \sim T_i/Z_ie$ is a flux function and $\varphi_1$ is found from the quasineutrality equation
\begin{equation}\label{eq:QN_totalenergyMoreBasic}
Z_i \int F_i\dd^3v = N_e.
\end{equation}
Here, $N_e$ is the electron density and $\int (\cdot)\dd^3 v \equiv \sum_\sigma \int (\cdot) B |v_{||}|^{-1} \dd\cE\dd\mu\dd\gamma$. To lowest order in $\sqrt{m_e/m_i}\ll
1$, only the adiabatic response of the electrons counts. Then,
\begin{equation}\label{eq:Ne}
N_e = n_e(\psi) \left\langle
\exp\left(\frac{e \varphi}{T_e(\psi)}\right)\right\rangle_\psi^{-1} \exp\left(\frac{e \varphi}{T_e(\psi)}\right)
,
\end{equation}
where $T_e$ is the electron temperature, $n_e$ is the flux-surface averaged electron density and $\langle \, \cdot \, \rangle_\psi$ denotes the flux-surface average operation, defined for a function $f(\psi,\alpha,l)$ as
\begin{equation}\label{eq:def_fluxsurfaceaverage}
\langle f \rangle_\psi = V'(\psi)^{-1}\int_0^{2\pi} \dd\alpha \int_0^{L(\psi,\alpha)}\dd l \, \Psi'_t B^{-1} f,
\end{equation}
where $V'(\psi)$ is the radial derivative of the volume enclosed by the flux surface labeled by $\psi$,
\begin{equation}\label{eq:def_volume}
V'(\psi) = \int_0^{2\pi} \dd\alpha \int_0^{L(\psi,\alpha)}\dd l \, \Psi'_t B^{-1}.
\end{equation}

In the quasineutrality equation defined by \eq{eq:QN_totalenergyMoreBasic} and \eq{eq:Ne} the expansions \eq{eq:expansionF1overnu} and \eq{eq:expansionvarphi1overnu} have not been employed yet. Using them, we obtain
\begin{eqnarray}\label{eq:QN_totalenergy}
\left(\frac{Z_i}{T_i}+\frac{1}{T_e}\right)\varphi_1 =
\frac{1}{en_i}\int
F_{i1} \, \dd^3v,
\end{eqnarray}
where we have assumed that $\varphi_1$ and the right side of \eq{eq:QN_totalenergy} have vanishing flux-surface average. The proof that this choice in the definition of $\varphi_1$ and $F_{i1}$ is possible is provided in reference \cite{Calvo13}.

From \eq{eq:QN_totalenergy}
and the fact that $F_{i1} \sim \nu_{i*}^{-1}\rho_{i*}F_{i0}^\cE$, one
obtains
\begin{equation}\label{eq:sizevarphi11overnu}
\varphi_1\sim\frac{\rho_{i*}}{\nu_{i*}}\varphi_0.
\end{equation}
This is the so-called $1/\nu$ regime~\cite{Kovrizhnykh84}, that exists for any stellarator (strictly speaking, for any stellarator that is not exactly omnigeneous).

The point that needs to be emphasized here is that the expansions \eq{eq:expansionF1overnu} and \eq{eq:expansionvarphi1overnu} do not work when $\nu_{i*} \lesssim \rho_{i*}$ because $F_{i1}$ becomes as large as
$F_{i0}^\cE$ and $\varphi_1$ becomes as large as $\varphi_0$ (see \eq{eq:sizeF11overnu} and \eq{eq:sizevarphi11overnu}). The regime $\nu_{i*} \lesssim \rho_{i*}$ is the subject of this paper, and we start to analyze it in the next subsection.

\subsection{Drift-kinetic equation when $\nu_{i*} \lesssim \rho_{i*}$ in stellarators close to
  omnigeneity}
\label{sec:DKEnustarMUCHSMALLERthanrhostar}

As explained above, the expansion of the distribution function and electrostatic potential employed in the $1/\nu$ regime (recall equations \eq{eq:expansionF1overnu}, \eq{eq:MaxwellianF0Etotal}, \eq{eq:sizeF11overnu}, \eq{eq:expansionvarphi1overnu} and \eq{eq:sizevarphi11overnu}), $\rho_{i*}\ll \nu_{i*}\ll 1$, is not valid when $\nu_{i*} \lesssim \rho_{i*}$. In order to understand what happens at collisionality values $\nu_{i*} \lesssim \rho_{i*}$ we go back to \eq{eq:DKEwithEtotal}, assume $\nu_{i*} \sim \rho_{i*}$ and expand in $\rho_{i*}$.

We take
\begin{equation}\label{eq:expansionFbelow1overnu}
  F_{i}
  = F_{i0}^\cE + F_{i1} + \dots
\end{equation}
with $F_{i1} \sim \rho_{i*} F_{i0}^\cE$. To lowest order in $\rho_{i*}$ equation \eq{eq:DKEwithEtotal} gives
\begin{equation}\label{eq:parallelstreamingequaltozero}
v_{||}\partial_l F_{i0}^\cE = 0.
\end{equation}
To solve \eq{eq:parallelstreamingequaltozero} and the next order equations, we employ a procedure similar to the one developed in \cite{Parra2014} for the $1/\nu$ regime. Equation \eq{eq:parallelstreamingequaltozero} implies that on an ergodic flux surface\footnote{
On a rational surface, passing particles follow periodic orbits and must be treated like standard trapped particles. Hence, there would be no splitting between $h_i$ and $g_i$ on a rational surface.} $F_{i0}^\cE$ can be written as
\begin{equation}\label{eq:F0constantalongtheline}
  F_{i0}^\cE = h_i(\psi,\cE,\mu,\sigma) + g_i(\psi,\alpha,\cE,\mu),
\end{equation}
where $g_i$ can be chosen such that it vanishes in the passing particle
region of phase space and $h_i$ cannot depend on $\sigma$ in the trapped particle region. In order to understand \eq{eq:F0constantalongtheline} observe, first, that the distribution function $F_{i0}^\cE$ cannot depend on $\alpha$ in the passing region of phase space because passing particles trace out a flux surface. Second, that in the trapped region of phase space $F_{i0}^\cE$ cannot depend on $\sigma$ because it has to be continuous at the bounce points. The split between $h_i$ and $g_i$ is defined up to a function independent of $\alpha$ that vanishes for passing particles. To completely determine $g_i$, we impose the condition
\begin{equation}\label{eq:condition_g1_integrationoveralpha}
\int_0^{2\pi} 
\dd\alpha \int_{l_{b_1}}^{l_{b_2}} \dd l \, \frac{g_i}{|v_{||}|}= 0. 
\end{equation}
There are other conditions that could be used to fix $g_i$.

The equation satisfied by $F_{i0}^\cE$ is found from averages of equation \eq{eq:DKEwithEtotal} to next order in $\rho_{i*}$. For passing 
particles one has to multiply the $O(\rho_{i*}v_{ti}L_0^{-1}F_{i0}^\cE)$ terms of \eq{eq:DKEwithEtotal} by $1/|v_{||}|$ and integrate over the flux 
surface, obtaining
\begin{eqnarray}\label{eq:DKEbelow1overnuNoExpansionpassing}  
\fl \int_0^{2\pi}
\dd\alpha \int_0^{L(\psi,\alpha)} \frac{1}{|v_{||}|} C_{ii}^{\cE}[F_{i0}^\cE,F_{i0}^\cE] \dd l = 0.
\end{eqnarray}
In order to get \eq{eq:DKEbelow1overnuNoExpansionpassing} we have employed $\partial_l F_{i0}^\cE = 0$, the fact that in the passing region $\partial_\alpha F_{i0}^\cE = 0$, and finally the property
\begin{eqnarray}
\fl \int_0^{2\pi}
\dd\alpha \int_0^{L(\psi,\alpha)} \frac{1}{|v_{||}|} (\bv_{M,i} + \bv_E)\cdot\nabla\psi
\dd l = 0
\end{eqnarray}
for passing trajectories.

For trapped particles we multiply the $O(\rho_{i*}v_{ti}L_0^{-1}F_{i0}^\cE)$ terms of \eq{eq:DKEwithEtotal} by $1/v_{||}$ and 
integrate over the orbit, arriving at
\begin{eqnarray}\label{eq:DKEbelow1overnuNoExpansiontrapped}  
\fl
-\partial_\psi J
\partial_\alpha F_{i0}^\cE
  + \partial_\alpha J \partial_\psi F_{i0}^\cE
 =
  \sum_\sigma
\frac{Z_ie\Psi'_t}{m_ic}\,
\int_{l_{b_1}}^{l_{b_2}}\frac{1}{|v_{||}|}
{C_{ii}^{\cE}[F_{i0}^\cE, F_{i0}^\cE]}\dd l.
\end{eqnarray}
Equation \eq{eq:DKEbelow1overnuNoExpansiontrapped} has conveniently been expressed 
in terms of the second adiabatic invariant
\begin{equation}\label{eq:defJ}
J(\psi,\alpha,\cE,\mu) := 2\int_{l_{b_1}}^{l_{b_2}}|v_{||}|\dd l
\end{equation}
by employing the
relations
\begin{eqnarray}\label{eq:DriftpsiIntermsofJ}
\fl  2\int_{l_{b_1}}^{l_{b_2}}\frac{1}{|v_{||}|}(\bv_{M,i}+\bv_{E})
\cdot\nabla\psi \, \dd l=
\frac{m_ic}{Z_ie\Psi'_t}\partial_\alpha J
\end{eqnarray}
and
\begin{eqnarray}\label{eq:DriftalphaIntermsofJ}
\fl  2\int_{l_{b_1}}^{l_{b_2}}\frac{1}{|v_{||}|}(\bv_{M,i}+\bv_{E})
\cdot\nabla\alpha \, \dd l=
-\frac{m_ic}{Z_ie\Psi'_t}\partial_\psi J,
\end{eqnarray}
that are derived in \ref{sec:relationJandDrifts}.

Given the profiles for ion density, ion temperature and radial electric field, the piece of the electrostatic potential that determines the tangential electric field is found from (recall \eq{eq:QN_totalenergyMoreBasic} and \eq{eq:Ne})
\begin{equation}\label{eq:QN_totalenergyNoDeltaExpansion}
Z_i \int F_{i0}^\cE \, \dd^3v = n_e(\psi) \left\langle
\exp\left(\frac{e \varphi}{T_e(\psi)}\right)\right\rangle_\psi^{-1} \exp\left(\frac{e \varphi}{T_e(\psi)}\right).
\end{equation}
In a generic stellarator one cannot go beyond \eq{eq:DKEbelow1overnuNoExpansionpassing}, \eq{eq:DKEbelow1overnuNoExpansiontrapped} and \eq{eq:QN_totalenergyNoDeltaExpansion}, that are a set of non-linear equations for the distribution function and the electrostatic potential. In particular, this means that without further assumptions, when $\nu_{i*} \lesssim \rho_{i*}$, one cannot deduce that $ F_{i0}^\cE$ be Maxwellian, and the drift-kinetic equation is clearly not radially local (note the term in \eq{eq:DKEbelow1overnuNoExpansiontrapped} containing $\partial_\psi F_{i0}^\cE$). However, we proceed to show that the situation is different if the stellarator is close to omnigeneity.

We take
\begin{equation}
\bB = \bB_0 + \delta \bB_1,
\end{equation}
where $\bB_0$ is omnigeneous, $\bB_0 \sim \bB_1$ and $0\le \delta \ll 1$, and assume that the expansion in $\delta$ is subsidiary with respect to the expansion in $\rho_{i*}$. As advanced in the Introduction, we only consider the case in which $B_1$ has small spatial derivatives tangent to the flux surface,
\begin{eqnarray}\label{eq:smallderivativeslandalpha}
\delta\partial_l B_1 \sim \delta L_0^{-1}B_0,\nonumber\\[5pt]
\delta\partial_\alpha B_1\sim \delta B_0.
\end{eqnarray}

The distribution function and the electrostatic potential are expanded as
\begin{equation}
F_{i0}^\cE = F_{i0}^{\cE (0)} + \delta F_{i0}^{\cE (1)} + \dots
\end{equation}
and
\begin{equation}
\varphi = \varphi_0(\psi) + \delta \varphi^{(1)} + \dots,
\end{equation}
where $F_{i0}^{\cE (1)} \sim F_{i0}^{\cE (0)} $ is the non-adiabatic correction of the distribution function and  $\varphi^{(1)} \sim \varphi_0$. We also expand $J$ as
\begin{equation}\label{eq:expansionJ}
J = J^{(0)} + \delta J^{(1)} + \dots,
\end{equation}
with
\begin{eqnarray}\label{eq:J0J1}
\fl  J^{(0)} = 2\int_{l_{b_{10}}}^{l_{b_{20}}}|v_{||}^{(0)}|\dd l,
  \nonumber\\[5pt]
\fl  J^{(1)} = -2
\int_{l_{b_{10}}}^{l_{b_{20}}}
\frac{1}{|v_{||}^{(0)}|}
\left(
\mu B_1(\alpha,l)
+\frac{Z_ie}{m_i}\varphi^{(1)}(\alpha,l)
\right)
\dd l.
\end{eqnarray}
Here $l_{b_{10}}$ and $l_{b_{20}}$ are the points that make
\begin{equation}\label{eq:vparzero}
\fl  v_{||}^{(0)}(\psi,\alpha,l,\cE,\mu) = \sigma \sqrt{2\left(
      \cE-\mu B_0(\psi,\alpha,l)-\frac{Z_ie}{m_i}
      \varphi_0(\psi)
    \right)}
\end{equation}
vanish and $J^{(0)}$ is independent of $\alpha$, which is the defining property of omnigeneity, as explained in Section \ref{sec:OmnigeneousStellarators}. The rigorous proof that the perturbation to $J^{(0)}$ is actually linear in $\delta$ when \eq{eq:smallderivativeslandalpha} is satisfied is contained in \cite{Parra16}. Finally, observe that we have assumed that $\varphi_0$ is a flux function. It can be proven that this follows from quasineutrality for an exactly omnigeneous magnetic field.

To lowest order in $\delta$ equation \eq{eq:DKEbelow1overnuNoExpansionpassing} gives
\begin{eqnarray}\label{eq:DKEbelow1overnuExpansionpassingLowestOrder}  
\fl \int_0^{2\pi}
\dd\alpha \int_0^{L^{(0)}(\psi)} \frac{1}{|v_{||}^{(0)}|} C_{ii}^{\cE (0)}[F_{i0}^{\cE (0)},F_{i0}^{\cE (0)}] \dd l = 0,
\end{eqnarray}
where the superindex $(0)$ in $C_{ii}^{\cE (0)}$ indicates that
  only $B_0$ has been kept in the kernel that defines the 
  collision operator. Analogously, $L^{(0)}(\psi)$ is the length of the magnetic field line for the omnigeneous configuration, and it has been stressed that it does not depend on $\alpha$.
  
  The lowest order terms of \eq{eq:DKEbelow1overnuNoExpansiontrapped} in the $\delta$ expansion are
\begin{eqnarray}\label{eq:DKEbelow1overnuExpansiontrappedLowestOrder}  
\fl
-\partial_\psi J^{(0)}
\partial_\alpha F_{i0}^{\cE (0)}
 =
  \sum_\sigma
\frac{Z_ie\Psi'_t}{m_ic}\,
\int_{l_{b_{10}}}^{l_{b_{20}}}\frac{1}{|v_{||}^{(0)}|}
{C_{ii}^{\cE (0)}[F_{i0}^{\cE (0)}, F_{i0}^{\cE (0)}]}\dd l,
\end{eqnarray}
where we have used $\partial_\alpha J^{(0)} = 0$ due to omnigeneity.

We solve equations \eq{eq:DKEbelow1overnuExpansionpassingLowestOrder} and \eq{eq:DKEbelow1overnuExpansiontrappedLowestOrder} by using the entropy production property of the collision operator. The lowest-order piece of \eq{eq:F0constantalongtheline} in the $\delta$ expansion
implies that $F_{i0}^{\cE (0)}$ does not depend on $\alpha$ in the passing region. Hence, we multiply \eq{eq:DKEbelow1overnuExpansionpassingLowestOrder}
by $-\ln F_{i0}^{\cE (0)}$ and find
\begin{eqnarray}\label{eq:DKEbelow1overnuExpansionpassingLowestOrderAux}  
\fl
-
\int_0^{2\pi}
\dd\alpha \int_0^{L^{(0)}(\psi)} \frac{1}{|v_{||}^{(0)}|} \ln F_{i0}^{\cE (0)} 
C_{ii}^{\cE (0)}[F_{i0}^{\cE (0)},F_{i0}^{\cE (0)}] \dd l = 0
\end{eqnarray}
in the passing region.

Similarly, we multiply  \eq{eq:DKEbelow1overnuExpansiontrappedLowestOrder}  by $-\ln F_{i0}^{\cE (0)}$, integrate the resulting expression over $\alpha$ and recall that omnigeneity implies that $\partial_\psi J^{(0)}$ does not depend on $\alpha$. We end up with
\begin{eqnarray}\label{eq:DKEbelow1overnuExpansiontrappedLowestOrderIntoveralpha}  
\fl
  -\sum_\sigma
\int_0^{2\pi}\dd\alpha \int_{l_{b_{10}}}^{l_{b_{20}}}\frac{1}{|v_{||}^{(0)}|}
\ln F_{i0}^{\cE (0)} {C_{ii}^{\cE (0)}[F_{i0}^{\cE (0)}, F_{i0}^{\cE (0)}]}\dd l
=
0
\end{eqnarray}
in the trapped region. Integrating \eq{eq:DKEbelow1overnuExpansionpassingLowestOrderAux} and \eq{eq:DKEbelow1overnuExpansiontrappedLowestOrderIntoveralpha} in velocity space and following an entropy-production argument, we deduce that $F_{i0}^{\cE (0)}$ is a Maxwellian distribution. Furthermore, it must have zero flow because $F_{i0}^{\cE (0)}$ cannot depend either on $l$ or on the gyrophase. Inserting the Maxwellian into  \eq{eq:DKEbelow1overnuExpansiontrappedLowestOrder}, we find that it is also independent of $\alpha$, leading to
\begin{equation}\label{eq:Fi0^(0)E}
F_{i0}^{\cE (0)}(\psi,\cE) = 
n_i(\psi)\left(\frac{m_i}{2\pi T_i(\psi)}\right)^{3/2}
\exp\left(-\frac{m_i\cE - Z_i e \varphi_0(\psi)}{T_i(\psi)}\right).
\end{equation}

We turn to the equations provided by terms that are linear in $\delta$ in \eq{eq:DKEbelow1overnuNoExpansionpassing} and \eq{eq:DKEbelow1overnuNoExpansiontrapped}. Using the decomposition \eq{eq:F0constantalongtheline}, we can write
\begin{equation}\label{eq:F^(1)constantalongtheline}
  F_{i0}^{\cE (1)} = h_i^{(1)}(\psi,\cE,\mu,\sigma) + g_i^{(1)}(\psi,\alpha,\cE,\mu),
\end{equation}
where $h_i^{(1)}$ cannot depend on $\sigma$ in the trapped particle region of phase space, and $g_i^{(1)}$ may be chosen such that it vanishes in the passing particle
region and such that
\begin{equation}\label{eq:BCgi(1)}
\int_0^{2\pi} g_i^{(1)}
\dd\alpha= 0. 
\end{equation}
Equation \eq{eq:BCgi(1)} is simply condition \eq{eq:condition_g1_integrationoveralpha} written to $O(\delta)$ by using \eq{eq:definitionomnigEquiv}.

To $O(\delta)$ equation \eq{eq:DKEbelow1overnuNoExpansionpassing} gives
\begin{eqnarray}\label{eq:DKEwithrhostar^2termsIntPassing}  
\fl \int_0^{2\pi}
\dd\alpha \int_0^{L^{(0)}(\psi)} \frac{1}{|v_{||}^{(0)}|} C_{ii}^{\cE,\ell (0)}[h_i^{(1)}] \dd l = 0,
\end{eqnarray}
where
\begin{equation}
C_{ii}^{\cE,\ell (0)}[h_i^{(1)}] = C_{ii}^{\cE (0)}[F_{i0}^{\cE (0)}, h_i^{(1)}] +
C_{ii}^{\cE (0)}[h_i^{(1)} , F_{i0}^{\cE (0)}] 
\end{equation}
is the linearization of the collision operator around $F_{i0}^{\cE (0)}$. In order to get \eq{eq:DKEwithrhostar^2termsIntPassing} we have employed that for passing trajectories
\begin{eqnarray}
\fl \int_0^{2\pi}
\dd\alpha \int_0^{L^{(0)}(\psi)} \frac{1}{|v_{||}^{(0)}|} C_{ii}^{\cE,\ell (0)}[g_i^{(1)}] \dd l = 0.
\end{eqnarray}
This is obtained by noting condition \eq{eq:BCgi(1)} and by using that, due to \eq{eq:definitionomnigEquiv}, $ \sum_\sigma
\int_{0}^{L^{(0)}(\psi)}
|v_{||}^{(0)}|^{-1}
{C_{ii}^{\cE,\ell (0)}[\, \cdot \,]} \dd l
$ is an operator whose coefficients are independent of $\alpha$ when acting on functions independent of 
$l$.
 
 The $O(\delta)$ terms of \eq{eq:DKEbelow1overnuNoExpansiontrapped} yield
  \begin{eqnarray}\label{eq:DKEwithrhostar^2termsIntTrapped}  
\fl
-\partial_\psi J^{(0)}
\partial_\alpha  F_{i0}^{\cE (1)}
  + \partial_\alpha J^{(1)} \Upsilon_i^\cE F_{i0}^{\cE (0)}
 =
  \sum_\sigma
\frac{Z_ie\Psi'_t}{m_ic}\,
\int_{l_{b_{10}}}^{l_{b_{20}}}\frac{\dd l}{|v_{||}^{(0)}|}
{C_{ii}^{\cE,\ell (0)}[F_{i0}^{\cE (1)}]},
\end{eqnarray}
where
\begin{equation}
\fl
\Upsilon_i^\cE = \frac{n_i'}{n_i} + \frac{T_i'}{T_i}
\left(
\frac{m_i(\cE - Z_i e\varphi_0/m_i)}{T_i}-\frac{3}{2}
\right)
+
\frac{Z_ie\varphi_0'}{T_i}.
\end{equation}

Next, we show that $h_i^{(1)}$ can be set equal to zero. We integrate \eq{eq:DKEwithrhostar^2termsIntTrapped} over $\alpha$, which gives
\begin{eqnarray}\label{eq:DKEwithrhostar^2termsIntTrappedIntegralAlpha}  
\fl
  \sum_\sigma
\int_0^{2\pi}\dd\alpha
\int_{l_{b_{10}}}^{l_{b_{20}}}\frac{1}{|v_{||}^{(0)}|}
{C_{ii}^{\cE,\ell (0)}[h_i^{(1)}]}
\dd l
=
0.
\end{eqnarray}
In order to obtain \eq{eq:DKEwithrhostar^2termsIntTrappedIntegralAlpha} we have used that in the trapped region
\begin{eqnarray}
\fl
  \sum_\sigma
\int_0^{2\pi}\dd\alpha
\int_{l_{b_{10}}}^{l_{b_{20}}}\frac{1}{|v_{||}^{(0)}|}
{C_{ii}^{\cE,\ell (0)}[g_i^{(1)}]}
\dd l
=
0.
\end{eqnarray}
This can be deduced by recalling \eq{eq:BCgi(1)} and by noting that, due to \eq{eq:definitionomnigEquiv}, $ \sum_\sigma
\int_{l_{b_{10}}}^{l_{b_{20}}}
|v_{||}^{(0)}|^{-1}
{C_{ii}^{\cE,\ell (0)}[\, \cdot \,]} \dd l
$ is an operator with coefficients independent of $\alpha$ when acting on functions independent of 
$l$. Multiplying \eq{eq:DKEwithrhostar^2termsIntPassing}  and 
\eq{eq:DKEwithrhostar^2termsIntTrappedIntegralAlpha} by $-h_i^{(1)}/F_{i0}^{\cE (0)}$, integrating over 
velocity space and applying again an entropy-production argument, we find that $h_i^{(1)}$ has to be a  
Maxwellian distribution with zero flow, and independent of $\alpha$ and $l$. Thus, it can be absorbed in the definition of $F_{i0}^{\cE (0)}$ and, 
from here on, we can assume
\begin{equation}
h_i^{(1)}\equiv 0
\end{equation}
without loss of generality.

Then, we only need to determine $g_i^{(1)}$, which is found from 
\eq{eq:DKEwithrhostar^2termsIntTrapped} by setting $h_i^{(1)}$ equal to zero. Namely,
\begin{eqnarray}\label{eq:DKEg}  
\fl
-\partial_\psi J^{(0)}
\partial_\alpha  g_i^{(1)}
  + \partial_\alpha J^{(1)} \Upsilon_i^\cE F_{i0}^{\cE (0)}
 =
  \sum_\sigma
\frac{Z_ie\Psi'_t}{m_ic}\,
\int_{l_{b_{10}}}^{l_{b_{20}}}\frac{\dd l}{|v_{||}^{(0)}|}
{C_{ii}^{\cE,\ell (0)}[g_i^{(1)}]}.
\end{eqnarray}
It is obvious, but still worth pointing out, that when $\rho_{i*} \ll
\nu_{i*}\ll 1$ the first term in \eq{eq:DKEg} can be
neglected and one recovers the equation for the dominant piece of the
distribution function in the $1/\nu$ regime of a stellarator close to
omnigeneity with a non-omnigeneous perturbation that has small gradients~\cite{Parra2014}.

Note that the orbit integrations in
\eq{eq:J0J1} and  \eq{eq:DKEg} only involve $B_0$ and $\varphi_0$. We use this fact to employ, in what follows, the coordinates
\begin{eqnarray}
v = \sqrt{2(\cE - Z_ie\varphi_0(\psi)/m_i)},
\nonumber\\[5pt]
\lambda = \frac{\mu}{\cE - Z_ie\varphi_0(\psi)/m_i},
\end{eqnarray}
in which the equations become simpler. We will not change the names of the functions $v_{||}^{(0)}$,
$\partial_\psi J^{(0)}$, $J^{(1)}$ and $g_i^{(1)}$ but
assume that they are expressed in coordinates $v$ and $\lambda$. Let
us be explicit to avoid any confusion. From here on, by $\partial_\psi
J^{(0)}$ and $J^{(1)}$ we understand
\begin{equation}\label{eq:dJ0/dpsi}
  \partial_\psi J^{(0)}
  =
  -\int_{l_{b_{10}}}^{l_{b_{20}}}
\frac{\lambda v\partial_\psi B_0 + 2Z_ie/(m_iv)\varphi'_0}
{\sqrt{1-\lambda B_0}}\dd l
\end{equation}
and
\begin{equation}\label{eq:dJ1/dalpha}
   J^{(1)}
  =
  -\int_{l_{b_{10}}}^{l_{b_{20}}}
\frac{
\lambda v B_1 + 2Z_ie/(m_iv)\varphi^{(1)}
}
{\sqrt{1-\lambda B_0}}\dd l.
\end{equation}

In these coordinates the
equation for $g_i^{(1)}$ reads, to the relevant order in $\delta$,
\begin{eqnarray}\label{eq:DKEtrappedOepsilong}
  \fl
-\partial_\psi J^{(0)}
\partial_\alpha  g_i^{(1)}
  + \partial_\alpha J^{(1)} \Upsilon_i F_{i0}
 =
  \sum_\sigma
\frac{Z_ie\Psi'_t}{m_ic}\,
\int_{l_{b_{10}}}^{l_{b_{20}}}\frac{\dd l}{|v_{||}^{(0)}|}
{C_{ii}^{\ell (0)}[g_i^{(1)}]},
\end{eqnarray}
where
\begin{equation}
\fl
F_{i0}(\psi,v) = 
n_i(\psi)\left(\frac{m_i}{2\pi T_i(\psi)}\right)^{3/2}
\exp\left(-\frac{m_iv^2}{2T_i(\psi)}\right),
\end{equation}
\begin{equation}
\fl
\Upsilon_i = \frac{n_i'}{n_i} + \frac{T_i'}{T_i}
\left(
\frac{m_iv^2}{2T_i}-\frac{3}{2}
\right)
+
\frac{Z_ie\varphi_0'}{T_i}
\end{equation}
and $C_{ii}^{\ell (0)}$ is the linearized collision operator corresponding to $B_0$ written
in coordinates $v$ and $\lambda$, given in \cite{Parra2014}.

Noting that $\int (\, \cdot \, )  \dd^2 S \equiv \int B^{-1}|\nabla\psi|(\, \cdot \, )  \dd\alpha \dd l$ and recalling \eq{eq:DriftpsiIntermsofJ}, the energy flux \eq{eq:defenergyfluxIntro} can be written as
\begin{eqnarray}\label{eq:GammaiLowColl}
\fl Q_i = \frac{\pi m_i^2c \delta^2}{Z_ie}
\int_0^\infty\dd v v^3
\left(
\frac{v^2}{2} + \frac{Z_i e \varphi_0}{m_i}
\right)
\int_{1/B_{0,{\rm max}}}^{1/B_{0,{\rm min}}}\dd\lambda \int_0^{2\pi}\dd\alpha\, 
\partial_\alpha J^{(1)} \, g_i^{(1)},
\end{eqnarray}
where $B_{0,{\rm min}}$ and $B_{0,{\rm max}}$ are the minimum and
maximum values of $B_0$ on the flux surface, respectively. The adiabatic response is absent from \eq{eq:GammaiLowColl} because its contribution vanishes (the same can be said about its contribution to the particle flux), as shown in \ref{sec:adiabaticresponsedoesnotcontribute}.

In particular, we have shown that $Q_i$ is proportional to $\delta^2$, the square of the size of the deviation from omnigeneity. In Section
\ref{sec:solutionDKEverylowcollisionality} we expand
\eq{eq:DKEtrappedOepsilong} for $\nu_{i*}\ll \rho_{i*}$ and give the expressions for \eq{eq:GammaiLowColl} in such collisionality regimes.

\section{Solution of the drift-kinetic equation \eq{eq:DKEtrappedOepsilong} when $\nu_{i*}\ll \rho_{i*}$}
\label{sec:solutionDKEverylowcollisionality}

Let us define the precession frequency due to the tangential drift
\begin{equation}
  \omega_\alpha (\psi,v,\lambda):= 
\frac{m_ic}{Z_ie\Psi'_t \tau_b^{(0)}}
\partial_\psi J^{(0)},
\end{equation}
where
\begin{equation}\label{eq:deftaub0}
\tau_b^{(0)}(\psi,v,\lambda) = \frac{2}{v}
\int_{l_{b_{10}}}^{l_{b_{20}}}\frac{\dd l}{\sqrt{1-\lambda B_0(\psi,\alpha,l)}}
\end{equation}
is the time that a particle trapped in $\bB_0$ takes to complete its
orbit. Note that $\tau_b^{(0)}$ does not depend on $\alpha$ due to
property \eq{eq:definitionomnigEquiv}, and therefore $\omega_\alpha$
is also independent of $\alpha$.

Typically, $\omega_\alpha\sim \rho_{i*}v_{ti}/L_0$, and equation
\eq{eq:DKEtrappedOepsilong} is solved by expanding in
$\nu_{ii}/\omega_\alpha~\sim \nu_{i*}/\rho_{i*}\ll 1$. We use the
notation
\begin{equation}\label{eq:gi(1)expansion}
 g_i^{(1)} = g_0 + g_1 + O((\nu_{ii}/\omega_\alpha)^2 F_{i0}),
\end{equation}
where $g_1/g_0 \sim O(\nu_{ii}/\omega_\alpha)$ and $g_0\sim
F_{i0}$.

To lowest order in the $\nu_{ii}/\omega_\alpha$ expansion, equation
\eq{eq:DKEtrappedOepsilong} gives
\begin{eqnarray}\label{eq:DKEsqrtnuZerothOrder}
  \fl
\partial_\alpha g_0
=
\frac{\partial_\alpha J^{(1)}}{\partial_\psi J^{(0)}}\Upsilon_i F_{i0}.
\end{eqnarray}
The solution of \eq{eq:DKEsqrtnuZerothOrder}, choosing
$\int_0^{2\pi}g_0 \, \dd\alpha = 0$, is
\begin{eqnarray}\label{eq:solutionDKEsqrtnuZerothOrder}
  \fl
g_0
=
\frac{1}{\partial_\psi J^{(0)}}\left(
J^{(1)}-\frac{1}{2\pi}\int_0^{2\pi}J^{(1)}\dd\alpha
\right)\Upsilon_i F_{i0}.
\end{eqnarray}

Using that
\begin{eqnarray}
\fl
\left(
J^{(1)}-\frac{1}{2\pi}\int_0^{2\pi}J^{(1)}\dd\alpha
\right) \partial_\alpha J^{(1)} = \frac{1}{2}\partial_\alpha
\left[
\left(
J^{(1)}-\frac{1}{2\pi}\int_0^{2\pi}J^{(1)}\dd\alpha
\right)^2
\right],
\end{eqnarray}
one proves that \eq{eq:solutionDKEsqrtnuZerothOrder}
does not contribute to \eq{eq:GammaiLowColl}. The next order terms of
\eq{eq:DKEtrappedOepsilong} in the $\nu_{ii}/\omega_\alpha$ expansion
yield
\begin{eqnarray}\label{eq:DKEsqrtnuFirstOrder}
  \fl
\partial_\alpha g_1=
-
\frac{1}{\omega_\alpha}
\overline{C_{ii}^{\ell (0)}[ g_0]},
\end{eqnarray}
where we have introduced a convenient notation for the orbit average,
\begin{equation}\label{eq:deforbitaverage}
\overline{(\cdot)} = \frac{1}{\tau_b^{(0)} v}\sum_\sigma
\int_{l_{b_{10}}}^{l_{b_{20}}}(\cdot)
\frac{\dd l}{\sqrt{1-\lambda B_0(\psi,\alpha,l)}}.
\end{equation}
Thus,
\begin{eqnarray}\label{eq:solutionDKEsqrtnuFirstOrder}
  \fl
g_1=
-
\frac{1}{\omega_\alpha}
\int^\alpha
\overline{C_{ii}^{\ell (0)}[ g_0]}\dd\alpha'
,
\end{eqnarray}
where the lower limit of the integral is selected so that
$\int_0^{2\pi}g_1 \dd\alpha = 0$.

When plugged into \eq{eq:GammaiLowColl}, this piece of
the distribution function gives a scaling
\begin{equation}\label{eq:GammaiOuter}
Q_i \sim \delta^2\frac{\nu_{ii}}{\omega_\alpha}\rho_{i*}n_iT_iv_{ti}S_\psi
\end{equation}
for the energy flux. However, this is not the
dominant contribution to $Q_i$. It turns out that the energy flux is
dominated by two small collisional layers that appear where \eq{eq:solutionDKEsqrtnuZerothOrder} is not a good approximation to $g_i^{(1)}$. This happens near the boundary between trapped and passing particles, and also near points where $\partial_\psi J^{(0)} = 0$. We study these layers in subsections \ref{sec:sqrtnuRegime} and
\ref{sec:superbananaplateauRegime}.

Finally, we advance a more subtle point. The necessity of solving the layers is not only tied to the calculation of transport fluxes. It is clear that one cannot say that the drift-kinetic equation has been completely solved unless $\varphi^{(1)}$ is known, because the latter enters the source term of the former via $\partial_\alpha J^{(1)}$. However, so far, $\varphi^{(1)}$ has not been found. It has to be determined from the quasineutrality equation \eq{eq:QN_totalenergyNoDeltaExpansion}. Expanding \eq{eq:QN_totalenergyNoDeltaExpansion} in
$\delta$ and choosing $\varphi^{(1)}$ such that it has vanishing flux-surface average, we obtain the equation that determines the tangential electric field
\begin{eqnarray}\label{eq:QNorderepsilon}
\left(\frac{Z_i}{T_i}+\frac{1}{T_e}\right)\varphi^{(1)} =
\frac{2\pi}{en_i} 
\int_0^\infty\dd v \int_{B^{-1}_{0,{\rm max}}}^{B^{-1}}\dd\lambda
\frac{v^3 B_0}{|v_{||}^{(0)}|}
g_i^{(1)}.
\end{eqnarray}
Here, we have used that in $\{v,\lambda,\sigma,\gamma\}$ coordinates 
$\int(\cdot)\dd^3v \equiv \sum_\sigma\int(\cdot) v^3 B / (2|v_{||}|)\dd v\dd\lambda\dd\gamma$ and that $g_i^{(1)}$ vanishes in the passing region, so that
the integral on the right side of \eq{eq:QNorderepsilon} is taken only
over trapped trajectories. Note that \eq{eq:QNorderepsilon} is consistent with the vanishing of the flux-surface average of $\varphi^{(1)}$ and with condition \eq{eq:BCgi(1)}.

We will prove later on (see subsection \ref{sec:quasineutrality_equation}) that the layer analyzed in subsection \ref{sec:superbananaplateauRegime} contributes to the quasineutrality equation as much as $g_0$, in general. Hence, to treat this layer, we need to calculate $\varphi^{(1)}$ self-consistently.

\subsection{Layer around the boundary between trapped and passing
  particles: the $\sqrt{\nu}$ regime}
\label{sec:sqrtnuRegime}

Recall that $ g_i^{(1)} \equiv 0$ in the passing region. The value of $\
g_i^{(1)}$ at the boundary of the trapped region is given
by $g_+ := g_0(\lambda_c)\neq 0$, with $\lambda_c = 1/B_{0,\max}$ and
$g_0$ given by \eq{eq:solutionDKEsqrtnuZerothOrder}\footnote{Sometimes, in order to ease the 
notation, we will omit
  some of the arguments of the functions. For example, in this section
  it will be common to display only the dependences on $\lambda$.}. Then, the
distribution function is not continuous. This discontinuity comes from
an incorrect treatment of the region around the interface between
passing and trapped particles. More specifically, it is the
consequence of dropping the collision term in that region. Usually,
this indicates~\cite{BenderOrszag} that there is a small layer in a
neighborhood of $\lambda_c$ where the distribution function develops
large variations in $\lambda$, and neglecting the collision term is
not correct. In the standard language of boundary-layer theory $g_0$
is the outer solution, and the inner solution of the boundary layer,
that we will denote by $\gbl$, remains to be found.

We have to replace
\eq{eq:gi(1)expansion} by
\begin{equation}\label{eq:gi(1)expansionlayer}
 g_i^{(1)} = g_0 + \gbl + \dots,
\end{equation}
where $\gbl$ satisfies the equation
\begin{eqnarray}\label{eq:boundarylayer}
  \fl
\omega_\alpha
\partial_\alpha  \gbl
+
\overline{C_{ii}^{\ell (0)}[\gbl]}= - \overline{C_{ii}^{\ell (0)}[g_0]}.
\end{eqnarray}
The collision operator acting on $g_0$ has been included on the right-hand
side of the previous equation because very close to $\lambda_c$ the
function $g_0$ varies fast with $\lambda$, and the right side of
\eq{eq:boundarylayer} actually diverges at $\lambda_c$, as we will see
below.

Equation
\eq{eq:boundarylayer} must be solved between $\lambda = \lambda_c$ and a value of $\lambda - \lambda_c > 0$ sufficiently large that $\gbl$ becomes small. Denote the width of the layer in the coordinate $\lambda$ by $\Delta_\lambda^{\sqrt{\nu}}\ll B_0^{-1}$ (its typical size is deduced below; see equation \eq{eq:Delta_lambda_sqrtnu}) and let $K$ be a constant that satisfies $K \gg 1$ and $K \Delta_\lambda^{\sqrt{\nu}}\ll B_0^{-1}$. Then, equation \eq{eq:boundarylayer} is viewed as an equation in the interval
$\lambda\in[\lambda_c, \lambda_K]$, with $\lambda_K - \lambda_c \sim K\Delta_\lambda^{\sqrt{\nu}}$. The boundary conditions are $\gbl(\lambda_c) = - g_+$ and $\gbl(\lambda_K) = 0$. At the end of this subsection we will conclude that the solution is asymptotically independent of $K$, as it should be.

Due to the boundary condition $\gbl(\lambda_c) = - g_+$, we know that $\gbl \sim g_0$. In addition, if the two terms on the left side of \eq{eq:boundarylayer} are to be
comparable in size, then the support of $\gbl$
(that is, the size of the boundary layer $\Delta_\lambda^{\sqrt{\nu}}$) has to be sufficiently small for the pitch angle 
scattering piece of the collision operator on the
left side of \eq{eq:boundarylayer},
\begin{equation}\label{eq:pitchAngleScattering}
C_{ii}^{\ell (0)}[\gbl] =
\frac{\nu_\lambda v_{||}^{(0)}}{v^2 B_0}\partial_\lambda\left(
v_{||}^{(0)}\lambda\partial_\lambda \gbl
\right) + \dots,
\end{equation}
to dominate. Here,
\begin{equation}
\fl
\nu_\lambda (v) = \frac{8\pi  n_i Z_i^4 e^4 \ln \Lambda}{m_i^2 v^3}
 \left[\mbox{erf} \left(\sqrt{m_i v^2 /(2T_i)}\right) - \chi \left(\sqrt{m_i v^2 /(2T_i)} \right)\right]
\end{equation}
is the pitch angle scattering frequency, $\ln \Lambda$ is the Coulomb
logarithm, $\chi (x) = [\mbox{erf} (x) - (2 x/\sqrt{\pi})
\exp(-x^2)]/(2x^2)$ and $\mbox{erf}(x) = (2/\sqrt{\pi}) \int_0^x \exp (
- t^2)\, \dd t$ is the error function.

In the boundary layer the pitch-angle scattering operator on the right side of 
\eq{eq:pitchAngleScattering} is dominated by the piece
that involves $\partial_\lambda^2 \gbl$; i.e. the term that contains $\partial_\lambda(
v_{||}^{(0)}\lambda)\partial_\lambda \gbl$ is small. The same happens for the
right side of \eq{eq:boundarylayer} close to $\lambda_c$, as will be
justified below. Therefore, \eq{eq:boundarylayer} can be approximated
by
\begin{eqnarray}\label{eq:boundarylayer2prev}
\fl
\omega_{\alpha}
\partial_\alpha  \gbl
+
\nu_\lambda\lambda\,
\overline{B_0^{-1} (1-\lambda B_0)}\,\partial^2_\lambda
\gbl
=
-\nu_\lambda\lambda\,
\overline{B_0^{-1} (1-\lambda B_0)}\,\partial^2_\lambda
g_0,
\end{eqnarray}
where again the coefficient multiplying $\partial_\lambda^2\gbl$ and
$\partial_\lambda^2 g_0$ does not depend on $\alpha$ due to
\eq{eq:definitionomnigEquiv}.

Due to the smallness of the boundary layer we can approximate this
equation further by taking $\lambda = \lambda_c$ in most terms; that is, equation 
\eq{eq:boundarylayer2prev} becomes
\begin{eqnarray}\label{eq:boundarylayer2}
\partial_\psi J^{(0)}  \partial_\alpha\gbl
+
\nu_\lambda\xi\partial^2_\lambda
\gbl
=
-\nu_\lambda\xi\partial^2_\lambda
g_0,
\end{eqnarray}
where
\begin{equation}
\fl
\xi(\psi,v) := \frac{Z_i e \Psi_t'}{m_i c}
  \frac{2\lambda_c}{v}\int_{l_{b_{10}}}^{l_{b_{20}}}B_0^{-1}
\sqrt{1-\lambda_c B_0(\psi,\alpha,l)}\,\dd l
.
\end{equation}
The dependence of $\partial_\psi J^{(0)}$ on $\lambda$ cannot be
neglected because $\partial_\psi J^{(0)}(\psi,v,\lambda)$ diverges
when $\lambda \to \lambda_c$. The point $\lambda = \lambda_c$ is a singular point
of the differential equation \eq{eq:boundarylayer2} and requires a
careful analysis. The right-hand side of \eq{eq:boundarylayer2} also
diverges at $\lambda_c$, as pointed out above. We proceed to
explain how these divergences emerge.

In \ref{sec:Expansionj0psi} we show that the asymptotic expansion of
$\partial_\psi J^{(0)}$ for small $\lambda - \lambda_c$ (with $\lambda
> \lambda_c$) is of the form\footnote{An identical calculation to the one carried out in \ref{sec:Expansionj0psi} for $\partial_\psi J^{(0)}$ shows that $\tau_b^{(0)}$ also diverges logarithmically when $\lambda\to\lambda_c$. This is not a problem in order to define the orbit-averaged drift-kinetic equation in the boundary layer because the number of particles for which $\nu_{ii}\tau_b^{(0)} \gg 1$ is exponentially small, $O(\exp(-1/\nu_{i*}))$, and therefore negligible in an asymptotic expansion in small $\nu_{i*}$.}
\begin{equation}\label{eq:expansionj0psi}
\partial_\psi J^{(0)} = 
a_1 \ln(B_{0,\max}(\lambda-\lambda_c)) 
+ a_2 + O(v_{ti}L_0 B_0 (\lambda-\lambda_c)/\psi),
\end{equation}
where
\begin{equation}\label{eq:a1}
  a_1 = \sqrt{\frac{1}{2\lambda_c}}
\sum_{k=1}^2
  \frac{\lambda_c v\partial_\psi B_0(l_{M,k})+ 2Z_ie/(m_iv)\partial_\psi\varphi_0}{\sqrt{|
\partial_l^2B_0(l_{M,k})|}}
\end{equation}
and the values $l_{M,k}$, for $k=1,2$, locate two consecutive absolute
maxima of $B_0$ when moving along the field line. 

The coefficient $a_2$ can be
computed from the relation
\begin{equation}\label{eq:a2}
a_2 = \lim_{\lambda\to\lambda_c}
\left(\partial_\psi J^{(0)} - a_1 \ln(B_{0,\max}(\lambda-\lambda_c))\right).
\end{equation}
Defining ${\tilde a_2}$ by the relation
\begin{equation}
a_2 = a_1 \ln\left(B_{0,\max}^{-1}{\tilde a_2}\right),
\end{equation}
one can recast \eq{eq:expansionj0psi} in the more convenient way
\begin{equation}\label{eq:expansionj0psi2}
\partial_\psi J^{(0)} = a_1 \ln({\tilde a_2}(\lambda-\lambda_c)) 
+ O(v_{ti}L_0 B_0(\lambda-\lambda_c)/\psi).
\end{equation}

Analogously, the asymptotic expansion of $
J^{(1)}$ yields
\begin{equation}\label{eq:expansionj1alpha}
 J^{(1)} = 
c_1 \ln(B_{0,\max}(\lambda-\lambda_c)) 
+ c_2 + O(v_{ti}L_0 B_0 (\lambda-\lambda_c)),
\end{equation}
where
\begin{equation}\label{eq:c1}
c_1 = \sqrt{\frac{1}{2\lambda_c}}
\sum_{k=1}^2
  \frac{\lambda_c v B_1(l_{M,k})+ 2Z_ie/(m_iv)\varphi_1(l_{M,k})}
{\sqrt{|\partial_l^2B_0(l_{M,k})|}}
\end{equation}
and
\begin{equation}\label{eq:c2}
c_2 = \lim_{\lambda\to\lambda_c}
\left(J^{(1)} - c_1 \ln(B_{0,\max}(\lambda-\lambda_c))\right).
\end{equation}
We rewrite \eq{eq:expansionj1alpha} as
\begin{equation}\label{eq:expansionj1alpha2}
J^{(1)} = c_1 \ln({\tilde a_2}(\lambda-\lambda_c)) +
{\tilde c}_2 + O(v_{ti}L_0 B_0(\lambda-\lambda_c)),
\end{equation}
with
\begin{equation}\label{eq:c2tilde}
{\tilde c}_2 = c_2 - c_1 \ln\left(B_{0,\max}^{-1}{\tilde a_2}\right).
\end{equation}

Using \eq{eq:expansionj0psi2} and \eq{eq:expansionj1alpha2}, equation  \eq{eq:boundarylayer2} becomes
\begin{eqnarray}\label{eq:boundarylayerInf}
\widehat{\partial_\psi J^{(0)}}  \partial_\alpha\gbl
+
\nu_\lambda\xi\partial^2_\lambda
\gbl
=
-\nu_\lambda\xi\partial^2_\lambda
\widehat{g_0},
\end{eqnarray}
where
\begin{equation}\label{eq:j0truncated}
\widehat{\partial_\psi J^{(0)}} = a_1 \ln({\tilde a_2}(\lambda-\lambda_c)),
\end{equation}
\begin{eqnarray}\label{eq:g0truncated}
\widehat{
g_0}
=
\frac{1}{\widehat{\partial_\psi J^{(0)}}}\left(
\widehat{J^{(1)}}-\frac{1}{2\pi}\int_0^{2\pi}\widehat{J^{(1)}}\dd\alpha
\right)\Upsilon_i F_{i0}
\end{eqnarray}
and
\begin{equation}\label{eq:j1truncated}
\widehat{J^{(1)}} = c_1 \ln({\tilde a_2}(\lambda-\lambda_c)) +
{\tilde c}_2.
\end{equation}
That is, \eq{eq:boundarylayerInf} is obtained from
\eq{eq:boundarylayer2} by keeping only the dominant terms in the
asymptotic expansions of $\partial_\psi J^{(0)}$ and $ J^{(1)}$ near
$\lambda_c$. It is clear from \eq{eq:j0truncated}, \eq{eq:g0truncated} and
\eq{eq:j1truncated} that the right side of \eq{eq:boundarylayerInf} 
diverges for $\lambda \to \lambda_c$. It is also obvious that, whereas both
$\partial_\psi J^{(0)}$ and $J^{(1)}$ diverge at
$\lambda_c$, $g_+:=g_0(\lambda_c)$ is finite, as it should be.

The solution of equation \eq{eq:boundarylayerInf} is more easily found
by first expanding $\gbl$ in Fourier modes with respect to the
coordinate $\alpha$. Define ${ g}_{{\rm bl},n}$, $g_{0,n}$ and ${
  g}_{+,n}$ by the relations
\begin{eqnarray}
\gbl(\alpha,\lambda) = \sum_{n=-\infty}^\infty { g}_{{\rm bl},n}(\lambda)
e^{in\alpha},\nonumber\\[5pt]
\widehat{g_0}(\alpha,\lambda) = \sum_{n=-\infty}^\infty { g}_{0, n}(\lambda) e^{in\alpha}
,\nonumber\\[5pt]
g_+(\alpha) = \sum_{n=-\infty}^\infty { g}_{+,n}e^{in\alpha}.
\end{eqnarray}
The Fourier coefficients ${ g}_{{\rm bl},0}$, ${ g}_{0, 0}$ and ${ g}_{+,0}$ are equal to zero because of condition \eq{eq:BCgi(1)}. Here, we have stressed the $\alpha$ and $\lambda$
dependence although $\gbl$, $\widehat{g_0}$ and $g_+$ also
depend on $\psi$ and $v$. Then, \eq{eq:boundarylayerInf} transforms into the set of ordinary
differential equations
\begin{eqnarray}\label{eq:boundarylayerInfFourier}
in \, \widehat{\partial_\psi J^{(0)}} { g}_{{\rm bl},n}
+
\nu_\lambda\xi\partial^2_\lambda
{ g}_{{\rm bl},n}
=
-\nu_\lambda\xi\partial^2_\lambda
{ g}_{0, n},
\end{eqnarray}
that must be solved with the boundary conditions
\begin{equation}\label{eq:BCFourier}
{ g}_{{\rm bl},n}(\lambda_c) = - {
  g}_{+,n}
\end{equation}
and
\begin{equation}\label{eq:BCFourierInfinity}
{ g}_{{\rm bl},n}(\lambda_K) = 0.
\end{equation}

At this point, we can explain why equation  \eq{eq:boundarylayerInfFourier} has not been extended up to $\lambda = \infty$ from the beginning. The reason is that there always exists a sufficiently large value of $\lambda$ such that the truncation $\widehat{\partial_\psi J^{(0)}}$ vanishes even though $\partial_\psi J^{(0)}$ may never vanish. We will see in subsection \ref{sec:superbananaplateauRegime} that points where $\partial_\psi J^{(0)} = 0$ correspond to another layer that, typically, gives non-negligible transport. To avoid points where $\widehat{\partial_\psi J^{(0)}} = 0$, we have imposed the boundary condition of \eq{eq:boundarylayerInfFourier} at a finite value of $\lambda_K$, with $\lambda_K - \lambda_c \sim K \Delta_\lambda^{\sqrt{\nu}}$. We must choose $K$ such that $\widehat{\partial_\psi J^{(0)}}$ does not vanish for $\lambda \in [\lambda_c,\lambda_K]$; i.e. such that $|\ln(\tilde{a_2}(\lambda - \lambda_c))| \ge |\ln(\tilde{a_2}(\lambda_K - \lambda_c))|$ for $\lambda \in [\lambda_c,\lambda_K]$.

The behavior of \eq{eq:boundarylayerInfFourier} in a vicinity of the singular point $\lambda = \lambda_c$ is analyzed in \ref{sec:IrregularSingularPoints}, where it is proven that the equation possesses solutions compatible with \eq{eq:BCFourier}.

We also expand $\widehat{J^{(1)}}$ in Fourier modes,
\begin{eqnarray}
\widehat{J^{(1)}}(\psi,\alpha,v,\lambda) = \sum_{n=-\infty}^\infty
\widehat{J^{(1)}}_n(\psi,v,\lambda) e^{in\alpha},
\end{eqnarray}
where
\begin{equation}\label{eq:modesofJ1hat}
\widehat{J^{(1)}}_n = c_{1,n} \ln({\tilde a}_2(\lambda-\lambda_c)) + {\tilde c}_{2,n}
\end{equation}
and where $c_{1,n}$ and ${\tilde c}_{2,n}$ are the coefficients of the Fourier expansions of $c_1$ and 
${\tilde c}_2$. Employing the solution for $\gbl$, we find that the contribution of the
boundary layer to the right side of \eq{eq:GammaiLowColl} is
\begin{eqnarray}\label{eq:EnergyFluxsqrtnu}
\fl Q_{i,\sqrt{\nu}} = -\frac{2\pi^2 m_i^2c\delta^2}{Z_ie}
\sum_{n=-\infty}^\infty
in
\int_0^\infty\dd v v^3
\left(
\frac{v^2}{2} + \frac{Z_i e \varphi_0}{m_i}
\right)
\int_{\lambda_c}^{\lambda_K}\dd\lambda\, 
\widehat{J^{(1)}}_{-n}
 \, { g}_{{\rm bl},n}.
\end{eqnarray}

We proceed to show the scaling of \eq{eq:EnergyFluxsqrtnu} with the square root of
the collisionality, to which the $\sqrt{\nu}$ regime owes its name. We will also prove that the scaling with the square root of the collisionality must actually be corrected by a logarithm due to the logarithmic singularities in \eq{eq:boundarylayerInfFourier}. Finally, we will show that the solution is independent of the constant $K$ as long as $|\ln(\tilde{a_2}(\lambda_K - \lambda_c))|$ is sufficiently large.

In \eq{eq:boundarylayerInfFourier}, we perform the change of coordinate
\begin{equation}\label{eq:changecoorlogcorr}
z = \sqrt{\frac{1}{\Delta^2}\ln \left(\frac{1}{{\tilde a}_2^2 \Delta^2 }\right)}\, (\lambda - \lambda_c),
\end{equation}
where
\begin{equation}
\Delta = \left(\frac{\nu_\lambda \xi}{|n a_1|}\right)^{1/2}.
\end{equation}
Then, equation \eq{eq:boundarylayerInfFourier} becomes
\begin{eqnarray}\label{eq:boundarylayerInfFourierChangeCoor}
\fl
i \frac{n a_1}{|n a_1|}
\left[
\frac{\ln z}
{
\ln \left(1 / ({\tilde a}_2 \Delta)^2\right)
}
-\frac{1}{2}
\left(
1
+
\frac{\ln[\ln (1 / ({\tilde a}_2 \Delta)^2)]}{\ln(1 / ({\tilde a}_2 \Delta)^2)}
\right)
\right]{ g}_{{\rm bl},n}
+
\partial_z^2 { g}_{{\rm bl},n} = \partial_z^2 g_{0,n},
\end{eqnarray}
with
\begin{eqnarray}\label{eq:g0nincoordinatez}
\fl g_{0,n}
=
\frac{1}{a_1}
\left[
c_{1,n} -
\frac{2\,{\tilde c}_{2,n}}{
\ln(1/({\tilde a}_2^2 \Delta^2))
+
\ln\left[
\ln(1/({\tilde a}_2 \Delta)^2))
\right] -2\ln z
}
\right]\Upsilon_i F_{i0}.
\end{eqnarray}

Employing that $z\sim 1$ in the layer and performing an expansion in the small quantity $1 / \ln (1 / ({\tilde a}_2 \Delta)^2) \ll 1$, one obtains to lowest order
\begin{eqnarray}\label{eq:boundarylayerInfFourierChangeCoorLowestOrder}
-\frac{i}{2}\frac{n a_1}{|n a_1|}
{ g}_{{\rm bl},n}
+
\partial_z^2 { g}_{{\rm bl},n}= 0,
\end{eqnarray}
where we have used, in particular, that the right side of \eq{eq:boundarylayerInfFourierChangeCoor} is small in $1 / \ln (1 / ({\tilde a}_2 \Delta)^2)$, as can be deduced by inspecting \eq{eq:g0nincoordinatez}. Equation \eq{eq:boundarylayerInfFourierChangeCoorLowestOrder} has an exponentially decaying solution with a characteristic width $\Delta_z\sim 1$. Using \eq{eq:changecoorlogcorr} to go back to the original coordinate $\lambda$, we find that the width of the layer in $\lambda$ is
\begin{equation}\label{eq:Delta_lambda_sqrtnu}
\Delta_\lambda^{\sqrt{\nu}} \sim
\left(\frac{\nu_\lambda \xi}{| n a_1 |}\right)^{1/2}
\frac{
1
}{\sqrt{\ln(1 / ({\tilde a}_2 \Delta)^2)}}.
\end{equation}
Therefore, not even to lowest order in $1 / \ln (1 / ({\tilde a}_2 \Delta)^2) \ll 1$ does the width of the layer scale exactly with the square root of the collision frequency. Although the logarithmic corrections do not change the
qualitative features of this collisionality regime, they must be accounted for in order to have accurate results for the
neoclassical fluxes. Noting the asymptotic expression \eq{eq:modesofJ1hat}, and using the change of coordinate \eq{eq:changecoorlogcorr} to rewrite the right side of \eq{eq:EnergyFluxsqrtnu}, we find that the 
size of $Q_{i,\sqrt{\nu}}$ is
\begin{equation}\label{eq:sizeQisqrtnuAgain}
Q_{i,\sqrt{\nu}}\sim \delta^2\frac{\nu_{ii}^{1/2}}{\omega_\alpha^{3/2}}
\sqrt{\ln(\omega_\alpha/\nu_{ii})}\rho_{i*}^2
n_iT_iv_{ti}^2 L_0^{-1}S_\psi.
\end{equation}
Finally, we point out that the expansion of \eq{eq:boundarylayerInfFourierChangeCoor} in the small quantity $1 / \ln (1 / ({\tilde a}_2 \Delta)^2)$ can be continued to higher orders. It is straightforward to check that, to any order, the boundary condition for large $z$ (equivalently, for large $\lambda$) can be imposed at $z= \infty$. In other words, the solution to equation \eq{eq:boundarylayerInfFourierChangeCoor} is independent of $K$ when $1 / \ln (1 / ({\tilde a}_2 \Delta)^2) \ll 1$.

Recall that $\varphi_1^{(1)}$ remains to be found. In order to write the precise form of the quasineutrality equation that determines $\varphi_1^{(1)}$ (given in subsection \ref{sec:quasineutrality_equation}), we have to solve for the layer around points where $\partial_\psi J^{(0)} = 0$. This is the subject of subsection \ref{sec:superbananaplateauRegime}.

\subsection{Layer around points where $\omega_\alpha = 0$: the
  superbanana-plateau regime}
\label{sec:superbananaplateauRegime}

The outer solution \eq{eq:solutionDKEsqrtnuZerothOrder} for the
distribution function is correct everywhere except near the boundary
between the passing and trapped regions (already treated in subsection
\ref{sec:sqrtnuRegime}) and in the neighborhood of points where
$\omega_\alpha = 0$ (equivalently, points where $\partial_\psi J^{(0)} = 0$). Around 
these `resonant points' the
$\nu_{ii}/\omega_\alpha \ll 1$ expansion is not valid. This region of
phase space is the subject of the present section.

In order to understand what happens in the vicinity of a point where
$\omega_\alpha = 0$, we go back to equation \eq{eq:DKEtrappedOepsilong}
and do not carry out the $\nu_{ii}/\omega_\alpha \ll 1$
expansion. That is, we consider the equation
\begin{eqnarray}\label{eq:DKEtrappedOepsilongSuperb}
\omega_\alpha
\partial_\alpha  g_i^{(1)}
 +
\overline{C_{ii}^{\ell (0)}[g_i^{(1)}]}
=
S
\end{eqnarray}
with
\begin{equation}\label{eq:defS}
  S(\psi,\alpha,v,\lambda) = 
 \frac{m_ic}{Z_ie\Psi'_t\tau_b^{(0)}}\partial_\alpha J^{(1)}
  \Upsilon_i F_{i0}.
\end{equation}
{Below we will find it useful to distinguish between the contributions to $\partial_\alpha J^{(1)}$ coming from $B_1$ and from $\varphi^{(1)}$. Defining
\begin{equation}\label{eq:J1_B}
   J_B^{(1)}
  =
  -\lambda v
  \int_{l_{b_{10}}}^{l_{b_{20}}}
\frac{
B_1
}
{\sqrt{1-\lambda B_0}}\dd l
\end{equation}
and
\begin{equation}\label{eq:J1_varphi}
   J_\varphi^{(1)}
  =
  -
  \frac{2Z_ie}{m_i v}
  \int_{l_{b_{10}}}^{l_{b_{20}}}
\frac{\varphi^{(1)}
}
{\sqrt{1-\lambda B_0}}\dd l,
\end{equation}
we write $S = S_B + S_\varphi$, where
\begin{equation}\label{eq:defS_B}
  S_B(\psi,\alpha,v,\lambda) = 
 \frac{m_ic}{Z_ie\Psi'_t\tau_b^{(0)}}\partial_\alpha J_B^{(1)}
  \Upsilon_i F_{i0}
\end{equation}
and
\begin{equation}
  S_\varphi(\psi,\alpha,v,\lambda) = 
 \frac{m_ic}{Z_ie\Psi'_t\tau_b^{(0)}}\partial_\alpha J_\varphi^{(1)}
  \Upsilon_i F_{i0}.
\end{equation}

We call $\lambda_r$ the values of $\lambda$ that satisfy $\omega_\alpha = 0$. Given an
omnigeneous magnetic field $\bB_0$, they are found from the equation (recall \eq{eq:dJ0/dpsi} and \eq{eq:deforbitaverage})
\begin{equation}\label{eq:condition_for_lambda_r}
\lambda_r \overline{\partial_\psi B_0}(\psi,\lambda_r)
=
-\frac{2Z_ie \varphi'_0(\psi)}
{m_i v^2}.
\end{equation}
Of course, in general $\lambda_r$ is a function of $\psi$ and $v$, $\lambda_r\equiv\lambda_r(\psi,v)$. The qualitative discussion on the number of zeroes of \eq{eq:condition_for_lambda_r} depends on the number of zeroes of \eq{eq:condition_for_lambda_r} for the particular case of $\varphi'_0 = 0$,
\begin{equation}\label{eq:condition_for_lambda_r_Erzero}
\overline{\partial_\psi B_0}(\psi,\lambda_{r0})
= 0.
\end{equation}
To fix ideas, we assume the common situation in which one, and only one value of $\lambda$ solves equation \eq{eq:condition_for_lambda_r_Erzero}. In this setting, for any value of $\varphi'_0$ and $v$, $\omega_\alpha$ vanishes at most for one value of $\lambda$. And for any given value of $\varphi'_0$, there exists a minimum value of $v$ such that $\omega_\alpha = 0$ for some value of $\lambda$. We denote this value of $v$ by $v_{\rm min}$.

Around $\lambda_r$,
\begin{equation}
  \omega_\alpha (\lambda) = 
  \partial_\lambda\omega_\alpha(\lambda_r)(\lambda-\lambda_r) 
+ O((\lambda-\lambda_r)^2),
\end{equation}
where the dependence on $\psi$ and $v$ has been
omitted for simplicity. The balance of the two terms on the left side
of \eq{eq:DKEtrappedOepsilongSuperb} implies that in a neighborhood of
$\lambda_r$ of size $\Delta_\lambda^{{\rm sb-p}}$,
\begin{equation}
\partial_\lambda\omega_\alpha(\lambda_r)\Delta_\lambda^{{\rm sb-p}}
\sim
\frac{\nu_{ii}}{B_0^2(\Delta_\lambda^{{\rm sb-p}})^2}.
\end{equation}
Since, typically, $\partial_\lambda\omega_\alpha(\lambda_r) \sim \rho_{i*}B_0
L_0^{-1} v_{ti}$, one finds
\begin{equation}\label{eq:sizelayersb-p}
B_0\Delta_\lambda^{{\rm sb-p}} \sim \left(\frac{\nu_{i*}}{\rho_{i*}}\right)^{1/3}\ll 1.
\end{equation}
In the particular case of a large-aspect-ratio tokamak with broken symmetry, this estimation coincides with the one obtained in \cite{Shaing2010}.

Denote by $g_{\rm rl}$ the distribution function in this `resonant layer'. The pitch-angle scattering
piece of the collision operator dominates the collision term in this layer,
\begin{equation}
\overline{C_{ii}^{\ell (0)}[g_{\rm rl}]} =
\overline{\frac{\nu_\lambda v_{||}^{(0)}}{v^2 B_0}\partial_\lambda\left(
v_{||}^{(0)}\lambda\partial_\lambda g_{\rm rl}
\right)}
 +\dots,
\end{equation}
and, in fact, we can keep only the term involving $\partial^2_\lambda
g_{\rm rl}$. Hence, in the resonant layer we write the drift
kinetic equation as
\begin{eqnarray}\label{eq:DKEresonantlayer}
\partial_\lambda \omega_{\alpha,r}(\lambda-\lambda_r)
\partial_\alpha  g_{\rm rl}
 +
\nu_\lambda\chi_r \partial_\lambda^2 g_{\rm rl}
=
S_{B, {r}} + \widehat{S_{\varphi}},
\end{eqnarray}
with
\begin{equation}
\chi_r(\psi,v) := \lambda_r
\overline{B_0^{-1} (1-\lambda_r B_0)
}\ \, ,
\end{equation}
\begin{equation}
\partial_\lambda \omega_{\alpha,r}(\psi,v)
:= \partial_\lambda\omega_\alpha(\psi,v,\lambda_r(\psi,v)),
\end{equation}
\begin{equation}
S_{B,r}(\psi,\alpha,v) := S_B(\psi,\alpha,v,\lambda_r(\psi,v))
\end{equation}
and
\begin{equation}
  \widehat{S_\varphi}(\psi,\alpha,v,\lambda) = 
 \frac{m_ic}{Z_ie\Psi'_t\tau_{b,{r}}^{(0)}}\partial_\alpha \widehat{J_\varphi^{(1)}}(\psi,\alpha,v,\lambda)
  \Upsilon_i F_{i0},
\end{equation}
with $\tau_{b,r}^{(0)} = \tau_{b}^{(0)}(\psi,v,\lambda_r(\psi,v))$ and
\begin{eqnarray}\label{eq:Jvarphi1truncated}
\fl
\widehat{J_\varphi^{(1)}}
  =
  -
  \frac{2Z_ie}{m_iv}
  \int_{\tilde l_{b_{10}}}^{l_R}
\frac{
\varphi^{(1)}
}
{\sqrt{\lambda_r |\partial_l B_0(l_L)|(l-l_L)
-(\lambda - \lambda_r) B_0(l_L)
}}\dd l
\nonumber\\[5pt]
\fl
\hspace{1cm}
 -
  \frac{2Z_ie}{m_iv}
  \int_{l_L}^{\tilde l_{b_{20}}}
\frac{
\varphi^{(1)}
}
{\sqrt{\lambda_r |\partial_l B_0(l_R)|(l_R-l)
-(\lambda - \lambda_r) B_0(l_R)
}}\dd l
\nonumber\\[5pt]
\fl
\hspace{1cm}
 -
  \frac{2Z_ie}{m_iv}
  \int_{l_L}^{l_R}\varphi^{(1)}
  \Bigg[
\frac{
1
}
{\sqrt{1-\lambda_r B_0(l)
}} - \frac{1}{\sqrt{\lambda_r |\partial_l B_0(l_L)|(l -l_L)
}}
\nonumber\\[5pt]
\fl
\hspace{1cm}
 - \frac{1}{\sqrt{\lambda_r |\partial_l B_0(l_R)|(l_R -l)
}}
\Bigg]
\dd l.
\end{eqnarray}
We have denoted by $l_L$ and $l_R$, respectively, the left and right bounce points of the orbit corresponding to $\lambda = \lambda_r$; i.e. the solutions for $l$ of $1-\lambda_r B_0(l) = 0$. In \eq{eq:Jvarphi1truncated}, $\tilde l_{{b_{10}}}$ and $\tilde l_{{b_{20}}}$ are approximations to the exact bounce points, $l_{{b_{10}}}$ and $l_{{b_{20}}}$, given by
\begin{equation}
\tilde l_{{b_{10}}} - l_L = \frac{B_0(l_L)}{\lambda_r |\partial_l B_0(l_L)|} (\lambda - \lambda_r)
\end{equation}
and
\begin{equation}
l_R - \tilde l_{{b_{20}}} = \frac{B_0(l_R)}{\lambda_r |\partial_l B_0(l_R)|} (\lambda - \lambda_r).
\end{equation}
Expression \eq{eq:Jvarphi1truncated} is an asymptotic approximation to $J_\varphi^{(1)}$ near the resonant value of the pitch-angle coordinate, as can be proven by using the techniques developed in \cite{Parra16}. Obviously, if $\varphi^{(1)}$ were regular everywhere, one could simply evaluate $J_\varphi^{(1)}$ at the resonant value $\lambda = \lambda_r$ (which would amount to retaining only the first term in square brackets in \eq{eq:Jvarphi1truncated}). However, in subsections \ref{sec:quasineutrality_equation} and \ref{sec:small_E_r_sb-p} we will show that if $\varphi'_0$ is small then $\varphi^{(1)}\sim (\rho_{i*} / \nu_{i*})^{1/6} T_i / e \gg T_i/e$ in a neighborhood of $l_L$ and $l_R$, and therefore the more elaborate asymptotic expression \eq{eq:Jvarphi1truncated} is nedeed.

Equation \eq{eq:DKEresonantlayer} is viewed as a differential equation in $\lambda\in(-\infty,\infty)$ with vanishing boundary conditions at infinity.  Note that a rescaling of the coordinate $\lambda$
\begin{equation}\label{eq:rescaling_sb-p}
x = \left(\frac{\nu_\lambda\chi_r}{\partial_\lambda \omega_{\alpha,r}}
\right)^{-1/3} (\lambda - \lambda_r)
\end{equation}
gives the expression
\begin{equation}\label{eq:Deltalambda_sb-p_better}
\Delta_\lambda^{{\rm sb-p}} =  \left(\frac{\nu_\lambda\chi_r}{\partial_\lambda \omega_{\alpha,r}}
\right)^{1/3}
\end{equation}
for the size of the layer needed to make the two terms on the left side of  \eq{eq:DKEresonantlayer} comparable. Then, the size of the distribution function in the layer can be estimated as\footnote{The perturbation to the Maxwellian has a size $\delta g_i^{(1)} \sim \delta g_{{\rm rl}}$ in the layer. From \eq{eq:sizeg_rl}, one might be worried that the perturbation to the Maxwellian becomes larger than the Maxwellian when $\nu_{i*} < \rho_{i*}\delta^3$. This is not a problem, however, because the analysis in this subsection does not apply to  such small values of the collisionality. This is explained in Section \ref{sec:Estimationnucritical} (see equation \eq{eq:nudelta_sb-p}).}
\begin{equation}\label{eq:sizeg_rl}
g_{\rm rl}
\sim \frac{S_{B,r}}{\partial_\lambda\omega_{\alpha,r}\Delta_\lambda^{{\rm sb-p}}}
\sim  \frac{1}{B_0 \Delta_\lambda^{{\rm sb-p}}} F_{i0}.
\end{equation}

Define the Fourier expansions
\begin{eqnarray}\label{eq:FourierexpansionsResLayer}
g_{\rm rl} = \sum_{n=-\infty}^\infty { g}_{{\rm rl},n}
e^{in\alpha},\nonumber\\[5pt]
S_{B, {r}} = \sum_{n=-\infty}^\infty ({S}_{B, {r}})_n e^{in\alpha},\nonumber\\[5pt]
\widehat{S_\varphi} = \sum_{n=-\infty}^\infty (\widehat{S_{\varphi}})_n e^{in\alpha}.
\end{eqnarray}
The coefficient ${ g}_{{\rm rl},0}$ vanishes due to \eq{eq:BCgi(1)}, and $({S}_{B, {r}})_0$ and $(\widehat{S_{\varphi}})_0$ vanish due to definition \eq{eq:defS}. Inserting
the expansions in \eq{eq:DKEresonantlayer} and noting that
$\partial_\lambda \omega_{\alpha,r}$ and $\chi_r$ do not depend on $\alpha$, we find an
ordinary differential equation for each mode ${ g}_{{\rm
    rl},n}$,
\begin{eqnarray}\label{eq:DKEresonantlayerFourier}
  in\partial_\lambda \omega_{\alpha,r}(\lambda-\lambda_r)
  { g}_{{\rm rl},n}
  +
  \nu_\lambda\chi_r \partial_\lambda^2
{ g}_{{\rm rl},n}
  =
  ({S}_{B, {r}})_n + (\widehat{S_{\varphi}})_n.
\end{eqnarray}
In terms of the solution to this set of equations, the energy flux~\eq{eq:GammaiLowColl} can be written as
\begin{eqnarray}\label{eq:Q_isb-p}
\fl Q_{i,{\rm sb-p}} = - \frac{2 \pi^2 m_i^2c \delta^2}{Z_ie}
\sum_{n=-\infty}^\infty in 
\int_{v_{\rm min}}^\infty
\dd v v^3
\left(
\frac{v^2}{2} + \frac{Z_i e \varphi_0}{m_i}
\right)
\int_{-\infty}^\infty
\dd\lambda
\Big[(J^{(1)}_{B, {r}})_{-n}
\nonumber\\[5pt]
+ (\widehat{J^{(1)}_\varphi})_{-n}
\Big] \, g_{{\rm rl},n}.
\end{eqnarray}
Here, $(\widehat{J^{(1)}_\varphi})_{n}$ are the coefficients of the Fourier expansion of $\widehat{J^{(1)}_\varphi}$ and 
$(J^{(1)}_{B, {r}})_{n}$ are the coefficients of the Fourier expansion of
\begin{equation}
J^{(1)}_{B, {r}}(\psi,\alpha,v) := J^{(1)}_B(\psi,\alpha,v,\lambda_r(\psi,v)).
\end{equation}

As long as $v_{\rm min} \lesssim v_{ti}$, the typical size of the energy flux is
\begin{eqnarray}\label{eq:EnergyFluxSuperbananaPlateau2size}
Q_{i,{\rm sb-p}} \sim \delta^2\rho_{i*}n_i T_i v_{ti}S_\psi,
\end{eqnarray}
which is a consequence of using \eq{eq:Deltalambda_sb-p_better}  and \eq{eq:sizeg_rl} in \eq{eq:Q_isb-p}. In particular, $Q_{i,{\rm sb-p}}$ does not scale with any power of collisionality. This is the most characteristic feature of the superbanana-plateau regime. Below, we explain that the estimates \eq{eq:Deltalambda_sb-p_better}, \eq{eq:sizeg_rl} and \eq{eq:EnergyFluxSuperbananaPlateau2size}, that are correct for sufficiently large radial electric field (see subsection \ref{sec:large_E_r_sb-p}), must be refined by including logarithmic corrections if the radial electric field is small enough (see subsection \ref{sec:small_E_r_sb-p}). The reason is that to obtain  \eq{eq:Deltalambda_sb-p_better}, \eq{eq:sizeg_rl} and \eq{eq:EnergyFluxSuperbananaPlateau2size}, we have skipped features of $\widehat{J^{(1)}_\varphi}$ that are important when $\varphi'_0$ is small.

Before turning to deal with the quasineutrality equation in the next subsection, it is useful to identify the piece of the distribution function $g_i^{(1)}$ out of the resonant layer. As we pointed out, $g_0$ diverges when $\partial_\psi J^{(0)} = 0$ and therefore it has to be replaced by $g_{\rm rl}$ in the layer. Sometimes, it is convenient to explicitly write $g_i^{(1)}$ as a sum of terms that are specifically associated to the layer and to the region external to the layer. This splitting is given by
\begin{equation}
g_i^{(1)} = g_0^{\rm out} + g_{\rm rl},
\end{equation}
where
\begin{eqnarray}\label{eq:def_g0_out}
  \fl
g_0^{\rm out}
=
g_0
-
\frac{1}{(\lambda - \lambda_r)\partial_\lambda\partial_\psi J^{(0)}(\lambda_r)}
\Bigg(
J_{B,r}^{(1)}-\frac{1}{2\pi}\int_0^{2\pi}J_{B,r}^{(1)}\dd\alpha
\nonumber\\[5pt]
\fl\hspace{1cm}
+
\widehat{J_\varphi^{(1)}}-\frac{1}{2\pi}\int_0^{2\pi}\widehat{J_\varphi^{(1)}}
\dd\alpha
\Bigg)\Upsilon_i F_{i0}.
\end{eqnarray}

\subsubsection{Quasineutrality equation.}
\label{sec:quasineutrality_equation}

We are ready to write more explicitly the quasineutrality equation \eq{eq:QNorderepsilon}, needed to find $\varphi^{(1)}$. The solution
\eq{eq:solutionDKEsqrtnuZerothOrder} does not contribute to transport
but it does contribute to \eq{eq:QNorderepsilon}. The component $\gbl$, associated to the $\sqrt{\nu}$ regime, gives a negligible contribution because $\gbl \sim F_{i0}$ and the size of the layer is small (see \eq{eq:Delta_lambda_sqrtnu}). However, in general, $g_{\rm rl}$ does contribute to  \eq{eq:QNorderepsilon} as much as $g_0$ (more precisely, as much as $g_0^{\rm out}$, defined in \eq{eq:def_g0_out}) due to the combination of \eq{eq:Deltalambda_sb-p_better}  and \eq{eq:sizeg_rl}.

Asymptotically, \eq{eq:QNorderepsilon} reads
\begin{eqnarray}\label{eq:QNorderepsilon2}
  \fl\left(\frac{Z_i}{T_i}+\frac{1}{T_e}\right)\varphi^{(1)} =
  \frac{2\pi}{en_i} 
  \int_0^\infty\dd v
  \int_{B^{-1}_{0,{\rm max}}}^{B^{-1}}\dd\lambda
  \frac{v^3 B_0}{|v_{||}^{(0)}|}
  g_0^{\rm out}
  \nonumber\\[5pt]
  \fl\hspace{1cm}
  +
  \frac{2\pi B_0}{en_i} 
  \int_{v_{\rm min}}^\infty
  \dd v\,
   v^2
\int_{-\infty}^{\lambda_L(l)}\dd\lambda
\frac{
g_{{\rm rl}}}
{\sqrt{\lambda_r |\partial_l B_0(l_L)|(l-l_L)
-(\lambda - \lambda_r) B_0(l_L)
}}
 \nonumber\\[5pt]
  \fl\hspace{1cm}
  +
  \frac{2\pi B_0}{en_i} 
  \int_{v_{\rm min}}^\infty
  \dd v\,
   v^2
\int_{-\infty}^{\lambda_R(l)}\dd\lambda
\frac{
g_{{\rm rl}}}
{\sqrt{\lambda_r |\partial_l B_0(l_R)|(l_R-l)
-(\lambda - \lambda_r) B_0(l_R)
}}
 \nonumber\\[5pt]
  \fl\hspace{1cm}
  +
  \frac{2\pi B_0}{en_i}
   \int_{v_{\rm min}}^\infty
  \dd v\,
   v^2
\Bigg[
\frac{
1
}
{\sqrt{1-\lambda_r B_0(l)
}} - \frac{1}{\sqrt{\lambda_r |\partial_l B_0(l_L)|(l -l_L)
}}
\nonumber\\[5pt]
\fl
\hspace{1cm}
 - \frac{1}{\sqrt{\lambda_r |\partial_l B_0(l_R)|(l_R -l)
}}
\Bigg]
\int_{-\infty}^\infty\dd\lambda \,
g_{{\rm rl}}
  .
\end{eqnarray}
Here,
\begin{equation}
\lambda_L(l) - \lambda_r = \frac{\lambda_r|\partial_l B_0(l_L)|(l-l_L)}{B_0(l_L)}
\end{equation}
and
\begin{equation}
\lambda_R(l) - \lambda_r = \frac{\lambda_r|\partial_l B_0(l_R)|(l_R-l)}{B_0(l_R)}.
\end{equation}
Of course, $\lambda_L$ and $\lambda_R$ depend on $\psi$ and $v$ as well, but for brevity we have only displayed the dependence on $l$, as we often do with other functions along the paper. The necessity for the complicated asymptotic expansion employed for the factor $|v_{||}^{(0)}|^{-1}$, instead of simply keeping the first term in square brackets in \eq{eq:QNorderepsilon2}, can be understood by observing that such a term diverges when $l = l_L$. Let us discuss this in more detail.

The function $\lambda_r(\psi,v)$ is obtained from condition \eq{eq:condition_for_lambda_r}. For the particular case of $\varphi'_0 = 0$, the resonant value of $\lambda$ is obtained from \eq{eq:condition_for_lambda_r_Erzero} and is denoted by $\lambda_{r0}(\psi)$, where we have stressed that $\lambda_{r0}$ does not depend on $v$. This will be important in what follows. The correction $\lambda_r - \lambda_{r0}$ that is linear in $\varphi'_0$ is found from
\begin{equation}\label{eq:condition_for_lambda_rE_rsmall}
\lambda_r - \lambda_{r0}
=
-
\frac{2Z_ie\varphi'_0(\psi)}{m_i v^2}
\Big[
\overline{\partial_\psi B_0}(\psi,\lambda_{r0}) +
\lambda_{r0}\partial_\lambda
\overline{\partial_\psi B_0
}(\psi,\lambda_{r0})
\Big]^{-1}.
\end{equation}
Defining $l_{L0}$ and $l_{R0}$ as the solutions for $l$ of $1-\lambda_{r0}B_0(l) = 0$, the corrections $l_L - l_{L0}$ and $l_R - l_{R0}$ are given by
\begin{equation}\label{eq:correctiontol_L0}
\fl
l_L - l_{L0}
=
- \frac{2Z_ie\varphi'_0(\psi)}{m_i v^2}\frac{B_0(l_{L0})}{\lambda_{r0}|\partial_l B_0(l_{L0})|}
\Big[
\overline{\partial_\psi B_0}(\psi,\lambda_{r0}) +
\lambda_{r0}\partial_\lambda
\overline{\partial_\psi B_0
}(\psi,\lambda_{r0})
\Big]^{-1},
\end{equation}
\begin{equation}\label{eq:correctiontol_R0}
\fl
l_R - l_{R0}
=
\frac{2Z_ie\varphi'_0(\psi)}{m_i v^2}
\frac{B_0(l_{R0})}{\lambda_{r0}|\partial_l B_0(l_{R0})|}
\Big[
\overline{\partial_\psi B_0}(\psi,\lambda_{r0}) +
\lambda_{r0}\partial_\lambda
\overline{\partial_\psi B_0
}(\psi,\lambda_{r0})
\Big]^{-1}.
\end{equation}

In order to make further progress we have to give an ordering for $\varphi'_0$, distinguishing the cases of small and large radial electric field as defined by the conditions
\begin{equation}\label{eq:def_smallE_r}
\lambda_r - \lambda_{r0} \sim \frac{2Z_i e \varphi'_0}{m_i v^2 \overline{\partial_\psi B_0}(\psi,\lambda_{r0}) } \ll \Delta_\lambda^{{\rm sb-p}}
\end{equation}
and
\begin{equation}\label{eq:def_largeE_r}
\lambda_r - \lambda_{r0} \sim \frac{2Z_i e \varphi'_0}{m_i v^2 \overline{\partial_\psi B_0}(\psi,\lambda_{r0}) } \gg \Delta_\lambda^{{\rm sb-p}},
\end{equation}
respectively.

\subsubsection{Small $\varphi'_0$.}
\label{sec:small_E_r_sb-p}

Let us take the first term that contains $g_{\rm rl}$ in \eq{eq:QNorderepsilon2}. Using \eq{eq:correctiontol_L0}, we can write
\begin{eqnarray}\label{eq:termtoillustrateLargeE_r}
 \fl
 \frac{2\pi B_0}{en_i} 
  \int_{v_{\rm min}}^\infty
  \dd v\,
   v^2
\int_{-\infty}^{\lambda_{L}(l)}\dd\lambda
\frac{
g_{{\rm rl}}}
{\sqrt{\lambda_{r} |\partial_l B_0(l_{L})|(l-l_{L})
-(\lambda - \lambda_{r}) B_0(l_{L})
}}\approx
\nonumber\\[5pt]
\fl
\hspace{0.5cm}
\frac{2\pi B_0}{en_i} 
  \int_{v_{\rm min}}^\infty
  \dd v\,
   v^2
\int_{-\infty}^{\lambda_{L}(l)}\dd\lambda
\frac{
g_{{\rm rl}}}
{\sqrt{\lambda_{r} |\partial_l B_0(l_{L})|(l-l_{L0} - \kappa_L)
-(\lambda - \lambda_{r}) B_0(l_{L})
}},
\end{eqnarray}
where
\begin{equation} \label{eq:defkappaL}
\fl
\kappa_L
=
- \frac{2Z_ie\varphi'_0(\psi)}{m_i v^2}\frac{B_0(l_{L0})}{\lambda_{r0}|\partial_l B_0(l_{L0})|}
\Big[
\overline{\partial_\psi B_0}(\psi,\lambda_{r0}) +
\lambda_{r0}\partial_\lambda
\overline{\partial_\psi B_0
}(\psi,\lambda_{r0})
\Big]^{-1}.
\end{equation}
If \eq{eq:def_smallE_r} holds, we can set $\varphi'_0 = 0$ in the previous expressions and find
\begin{eqnarray}\label{eq:termtoillustrate}
 \fl
 \frac{2\pi B_0}{en_i} 
  \int_{v_{\rm min}}^\infty
  \dd v\,
   v^2
\int_{-\infty}^{\lambda_{L}(l)}\dd\lambda
\frac{
g_{{\rm rl}}}
{\sqrt{\lambda_{r} |\partial_l B_0(l_{L})|(l-l_{L})
-(\lambda - \lambda_{r}) B_0(l_{L})
}} = 
\nonumber\\[5pt]
\fl \hspace{0.5cm} \frac{2\pi B_0}{en_i} 
  \int_{0}^\infty
  \dd v\,
   v^2
\int_{-\infty}^{\lambda_{L0}(l)}\dd\lambda
\frac{
g_{{\rm rl}}}
{\sqrt{\lambda_{r0} |\partial_l B_0(l_{L0})|(l-l_{L0})
-(\lambda - \lambda_{r0}) B_0(l_{L0})
}},
\end{eqnarray}
where $\lambda_{L0}$ stands for the function $\lambda_L$ in the particular case of $\varphi'_0 = 0$. The key observation is that the quantity under the square root on the right side of the last equation is independent of $v$. Then, at $l = l_{L0}$, the right side of \eq{eq:termtoillustrate} becomes
\begin{eqnarray}\label{eq:termtoillustrate2}
 \fl
 \frac{2\pi B_0}{en_i} 
  \int_{0}^\infty
  \dd v\,
   v^2
\int_{-\infty}^{\lambda_{r0}}\dd\lambda
\frac{
g_{{\rm rl}}}
{\sqrt{
(\lambda_{r0} - \lambda) B_0(l_{L0})
}}.
\end{eqnarray}
Therefore, when $\varphi'_0$ is small, the first term that contains $g_{\rm rl}$ in \eq{eq:QNorderepsilon2} gives a contribution to $\varphi^{(1)}$ whose typical size is
\begin{eqnarray}
\fl
\varphi^{(1)} \sim
\frac{m_i v_{ti}^5}{e n_i} B_0 \Delta_\lambda^{{\rm sb-p}} g_{\rm rl} \frac{1}{(B_0 \Delta_\lambda^{{\rm sb-p}})^{1/2}} \quad\mbox{ when } l - l_{L0} \sim B_0 \Delta_\lambda^{{\rm sb-p}} L_0
\end{eqnarray}
and
\begin{eqnarray}
\fl
\varphi^{(1)} \sim
\frac{m_i v_{ti}^5}{e n_i} B_0 \Delta_\lambda^{{\rm sb-p}} g_{\rm rl} \frac{1}{\sqrt{(l - l_{L0})/L_0}} \quad\mbox{ when } l - l_{L0} \gg B_0 \Delta_\lambda^{{\rm sb-p}} L_0.
\end{eqnarray}
When inserted in \eq{eq:Jvarphi1truncated}, this piece of $\varphi^{(1)}$ gives a contribution to 
$\widehat{J_\varphi^{(1)}}$ that scales as
\begin{equation}\label{eq:sizewidehatJvarphi}
\widehat{J_\varphi^{(1)}} \sim 
\frac{v_{ti}^4 L_0}{ n_i} B_0 \Delta_\lambda^{{\rm sb-p}} g_{\rm rl} \ln(B_0 \Delta_\lambda^{{\rm sb-p}}).
\end{equation}

We are ready to show why the estimates \eq{eq:Deltalambda_sb-p_better}, \eq{eq:sizeg_rl} and \eq{eq:EnergyFluxSuperbananaPlateau2size} are not completely correct if $\varphi'_0$ is small. Inspecting the size of each term in \eq{eq:DKEresonantlayer} and recalling \eq{eq:sizewidehatJvarphi}, one concludes that the width of the layer, $\Delta_\lambda^{{\rm sb-p}}$, is determined by balancing the collision term and the last term on the right side of equation \eq{eq:DKEresonantlayer}. The result is
\begin{equation}
B_0 \Delta_\lambda^{{\rm sb-p}} \sim \left( \frac{\nu_{i*}}{\rho_{i*}} \right)^{1/3} \left[
\ln(\rho_{i*}/\nu_{i*})
\right]^{-1/3}.
\end{equation}
The size of the distribution function is found by balancing the two terms on the right side of \eq{eq:DKEresonantlayer}, obtaining
\begin{equation}
g_{\rm rl} \sim \left(\frac{\rho_{i*}}{\nu_{i*}}\right)^{1/3}
\left[
\ln(\rho_{i*}/\nu_{i*})
\right]^{-2/3}F_{i0}.
\end{equation}
Then, the ion energy flux \eq{eq:Q_isb-p} scales as
\begin{eqnarray}
Q_{i,{\rm sb-p}} \sim \delta^2
\frac{\rho_{i*}}{
\ln(\rho_{i*}/\nu_{i*})
}
n_i T_i v_{ti}S_\psi.
\end{eqnarray}

\subsubsection{Large $\varphi'_0$.}
\label{sec:large_E_r_sb-p}

We consider again the first term that contains $g_{\rm rl}$ in \eq{eq:QNorderepsilon2} and recall expressions \eq{eq:termtoillustrateLargeE_r} and \eq{eq:defkappaL}. Now, assume that \eq{eq:def_largeE_r} holds. Then, $\lambda_r |\partial_l B_0(l_L) \kappa_L| \gg |\lambda - \lambda_r| B_0(l_L)$, and we can neglect 
$(\lambda - \lambda_r)B_0(l_L)$ in the quantity under the square root in \eq{eq:termtoillustrateLargeE_r}. The same argument can be applied to the second term containing $g_{\rm rl}$ in \eq{eq:QNorderepsilon2}. Therefore, if
\eq{eq:def_largeE_r} is satisfied, \eq{eq:QNorderepsilon2} simplifies to
\begin{eqnarray}\label{eq:QNorderepsilon2_Large_E_r}
  \fl\left(\frac{Z_i}{T_i}+\frac{1}{T_e}\right)\varphi^{(1)} =
  \frac{2\pi}{en_i} 
  \int_0^\infty\dd v
  \int_{B^{-1}_{0,{\rm max}}}^{B^{-1}}\dd\lambda
  \frac{v^3 B_0}{|v_{||}^{(0)}|}
  g_0^{\rm out}
  \nonumber\\[5pt]
  \fl\hspace{1cm}
  +
  \frac{2\pi B_0}{en_i}
   \int_{v_{\rm min}}^\infty
  \dd v\,
\frac{
v^2
}
{\sqrt{1-\lambda_r B_0(l)
}} 
\int_{-\infty}^\infty\dd\lambda \,
g_{{\rm rl}}
\end{eqnarray}
and $\widehat{J_\varphi^{(1)}}$, defined in \eq{eq:Jvarphi1truncated}, can be simply replaced by $J_{\varphi,r}^{(1)}$, where
\begin{eqnarray}\label{eq:Jvarphi1truncated_Er_large}
\fl
J_{\varphi,r}^{(1)}
  =
  -
  \frac{2Z_ie}{m_iv}
  \int_{l_L}^{l_R}
\frac{
\varphi^{(1)}
}
{\sqrt{1-\lambda_r B_0(l)
}} 
\dd l.
\end{eqnarray}

When $\varphi'_0$ satisfies \eq{eq:def_largeE_r}, we can solve \eq{eq:DKEresonantlayerFourier} analytically. Its solution vanishing at infinity is
\begin{equation}\label{eq:solmodesgresonantlayer}
{ g}_{{\rm rl},n}
=
-\frac{{ S}_{r,n}}{\partial_\lambda \omega_{\alpha,r} n^{2/3}\lambda_r\beta}
\int_0^\infty
\exp\left(
i\frac{n^{1/3}}{\beta}\frac{\lambda-\lambda_r}{\lambda_r}z-\frac{1}{3}z^3
\right)\dd z,
\end{equation}
where we have defined
\begin{equation}\label{eq:defbeta}
\beta := \left(\frac{\nu_\lambda\chi_r}{\partial_\lambda \omega_{\alpha,r}\lambda_r^3}
\right)^{1/3}\ll 1,
\end{equation}
$S_{r,n}$ are the coefficients of the Fourier expansion of $S_r$,
\begin{equation}
S_r(\psi,\alpha,v) := S(\psi,\alpha,v,\lambda_r(\psi,v)),
\end{equation}
and $S = S_B + S_\varphi$ has been defined in \eq{eq:defS}. Note that for $S_r$ to be well defined, it is essential that \eq{eq:Jvarphi1truncated_Er_large} be correct as the asymptotic expression of $J_\varphi^{(1)}$ near $\lambda = \lambda_r$, and this is only true as long as  condition \eq{eq:def_largeE_r} is met. Then, the contribution of resonant particles to
\eq{eq:GammaiLowColl} is
\begin{eqnarray}\label{eq:EnergyFluxSuperbananaPlateau}
\fl Q_{i,{\rm sb-p}} = -\frac{2\pi^2 m_i^2c\delta^2}{Z_ie}
\sum_{n=-\infty}^\infty in \int_{v_{\rm min}}^{\infty}
\dd v v^3
\left(
\frac{v^2}{2} + \frac{Z_i e \varphi_0}{m_i}
\right)
\int_{-\infty}^\infty\dd\lambda\, 
J^{(1)}_{r,-n} \, { g}_{{\rm rl},n}
\nonumber\\[5pt]
\hspace{0.5cm}
\fl=-
\frac{4\pi^2 m_i^3c^2\delta^2}{Z_i^2e^2 \Psi'_t}
\sum_{n=1}^\infty\int_{v_{\rm min}}^{\infty}\dd v v^3
\left(
\frac{v^2}{2} + \frac{Z_i e \varphi_0}{m_i}
\right)
\frac{n^{4/3}}{\partial_\lambda \omega_{\alpha,r}\tau_{b,r}^{(0)}\lambda_r}
 \Upsilon_i F_{i0}
\int_{-\infty}^\infty\dd\lambda\, 
|J^{(1)}_{r,n}|^2\Bigg\{
\nonumber\\[5pt]
\hspace{0.5cm}
\fl
\frac{
1}{\beta}
\int_0^\infty
\cos
\left(
\frac{n^{1/3}}{\beta}\frac{\lambda-\lambda_r}{\lambda_r}z
\right)
\exp\left(
-\frac{1}{3}z^3
\right)\dd z
\Bigg\},
\end{eqnarray}
where we have defined
\begin{equation}
J^{(1)}_r(\psi,\alpha,v) = J^{(1)}(\psi,\alpha,v,\lambda_r(\psi,v))
\end{equation}
and
\begin{eqnarray}
J^{(1)}_r(\psi,\alpha,v) = \sum_{n=-\infty}^\infty
 J^{(1)}_{r,n}(\psi,v) e^{in\alpha}.
\end{eqnarray}

Next, we prove that the right side of
\eq{eq:EnergyFluxSuperbananaPlateau} has a non-zero limit when $\beta
\to 0$. For this, we employ the identity
\begin{equation}\label{eq:DiracdeltaRepresentation}
\lim_{\beta\to 0}
\frac{1}{\beta}
\int_0^{\infty}
e^{-z^3/3}\cos\left(\frac{1}{\beta} x z\right)\dd z = \pi\delta(x)
\end{equation}
and the property $\delta(ax) = |a|^{-1}\delta(x)$, where
$\delta(\cdot)$ is the Dirac delta distribution and $a$ is a real
number. Then, for $\beta \ll 1$, the asymptotically dominant term is
\begin{eqnarray}\label{eq:EnergyFluxSuperbananaPlateau2}
\fl Q_{i,{\rm sb-p}} = 
-
\frac{4\pi^3 m_i^3c^2\delta^2}{Z_i^2e^2 \Psi'_t}
\sum_{n=1}^\infty
n
\int_{v_{\rm min}}^{\infty}\dd v v^3
\left(
\frac{v^2}{2} + \frac{Z_i e \varphi_0}{m_i}
\right)
\frac{1}{\partial_\lambda \omega_{\alpha,r}  \tau_{b,r}^{(0)}}
 \Upsilon_i F_{i0}
|J^{(1)}_n|^2.
\end{eqnarray}

\subsection{Formula for the ion energy flux when $\nu_{i*}\ll \rho_{i*}$}
\label{sec:additive formula}

Since the layers studied in subsections \ref{sec:sqrtnuRegime} and
\ref{sec:superbananaplateauRegime} are small and, in general, they are located
around different points of phase space, their
contributions to transport are additive. This means that we can write,
for $\nu_{i*}\ll \rho_{i*}$,
\begin{equation}\label{eq:Qiadditiveformula}
Q_i = Q_{i,\sqrt{\nu}} + Q_{i,{\rm sb-p}},
\end{equation}
where $Q_{i,\sqrt{\nu}}$ is given by \eq{eq:EnergyFluxsqrtnu} and
$Q_{i,{\rm sb-p}}$ is given by
\eq{eq:Q_isb-p}. The weight of each term in
\eq{eq:Qiadditiveformula} is determined by the value of $v_{\rm
  min}$. Typically, the estimate \eq{eq:EnergyFluxSuperbananaPlateau2size} will be supressed by a 
  factor $\exp(-v_{\rm min}^2/v_{ti}^2)$. Recalling also the estimate \eq{eq:sizeQisqrtnuAgain}, we deduce that  the
superbanana-plateau regime dominates over the
$\sqrt{\nu}$ regime when
\begin{equation}
\frac{v_{\rm min}}{v_{ti}} \ll \sqrt{\ln\left(\frac{\omega_\alpha}{\nu_{ii}}\right)}\,.
\end{equation}
Conversely, the
$\sqrt{\nu}$ regime dominates over the superbanana-plateau regime when
\begin{equation}
\frac{v_{\rm min}}{v_{ti}} \gg \sqrt{\ln\left(\frac{\omega_\alpha}{\nu_{ii}}\right)}\,.
\end{equation}

Finally, we note that the value of $v_{\rm min}$ is set by the
size of $\varphi'_0$, but also by the specific
$\lambda$-dependence of $\overline{\partial_\psi B_0}(\psi,\lambda_{r0})$ (recall condition \eq{eq:condition_for_lambda_r_Erzero}).

\section{Calculation of the radial electric field}
\label{sec:varphi1andvarphi0}

The radial electric field, determined by $\varphi_0'$, is one of the
quantities that are routinely computed in standard neoclassical
calculations. It is found by imposing that the radial electric current
vanish.

Let us denote by $\Gamma_i$ and $\Gamma_e$ the radial fluxes of ions
and electrons. The radial electric field is obtained by imposing
\begin{equation}
Z_i e \Gamma_i - e \Gamma_e = 0.
\end{equation}
To lowest order in a mass ratio expansion $\sqrt{m_e/m_i}\ll 1$
this is equivalent to the condition
\begin{equation}\label{eq:Gammai0}
\Gamma_i = 0.
\end{equation}

The calculation of $\Gamma_i$ is completely analogous to that of
$Q_i$. Hence, asymptotically, \eq{eq:Gammai0} amounts to the condition
\begin{equation}\label{eq:Gammaiadditiveformula}
\Gamma_{i,\sqrt{\nu}} + \Gamma_{i,{\rm sb-p}} = 0,
\end{equation}
where
\begin{eqnarray}\label{eq:Gamma_i_sqrtnu_secE_r}
\fl \Gamma_{i,\sqrt{\nu}} = -\frac{2\pi^2 m_i c\delta^2}{Z_ie}
\sum_{n=-\infty}^\infty
in
\int_0^\infty\dd v v^3
\int_{\lambda_c}^{\lambda_K}\dd\lambda\, 
\widehat{J^{(1)}}_{-n}
 \, { g}_{{\rm bl},n}
\end{eqnarray}
and
\begin{eqnarray}\label{eq:Gamma_i_sb-p_sec_E_r}
\fl \Gamma_{i,{\rm sb-p}} = - \frac{2 \pi^2 m_i c \delta^2}{Z_ie}
\sum_{n=-\infty}^\infty in 
\int_{v_{\rm min}}^{\infty}
\dd v v^3
\int_{-\infty}^\infty
\dd\lambda
\Big[(J^{(1)}_{B, {r}})_{-n}
+ (J^{(1)}_\varphi)_{-n}
\Big] \, g_{{\rm rl},n}.
\end{eqnarray}

\section{Estimation of  $\nu_{\delta *}$}
\label{sec:Estimationnucritical}

In Section \ref{sec:solutionDKEverylowcollisionality} we have solved
the drift-kinetic equation and computed $Q_i$ when $\nu_{i*}\ll
\rho_{i*}$. But we have advanced in the Introduction that our results
are not valid for arbitrarily small $\nu_{i*}$. There exists a value of
the collisionality, that we call $\nu_{\delta *}$, below which equation
\eq{eq:Qiadditiveformula} is expected to be incorrect because the approximation to the
drift-kinetic equation in \eq{eq:DKEtrappedOepsilong} is incorrect.
Hence, it is more precise to say that our results in Section \ref{sec:solutionDKEverylowcollisionality} are correct when
$\nu_{\delta *} \ll \nu_{i*}\ll \rho_{i*}$. In this section we explain the
reason for the existence of $\nu_{\delta *}$ and estimate its value.

The limitations of  equation \eq{eq:DKEtrappedOepsilong} for sufficiently small
$\nu_{i*}$ are well understood by inspecting the drift-kinetic equation
written with the parallel velocity $u$ and the magnetic moment $\mu$ as independent coordinates.  The drift-kinetic equation in terms of these coordinates is calculated in \cite{Calvo13} to second order in a $\rho_{i*}$ expansion. If we denote by $\check F_{i}(\bR,u,\mu)$ the distribution function expressed in coordinates $\{\bR, u, \mu,\gamma\}$, and by $\check F_{i1}(\bR,u,\mu) \sim \delta \check F_{i}$ the deviation of $\check F_{i}(\bR,u,\mu)$ from a Maxwellian distribution, we can check that the drift-kinetic equation in \cite{Calvo13} contains a term 
of the form
\begin{equation}
u\kappabf\cdot (\bv_{\nabla B,i} + \bv_{E,0})\partial_u \check F_{i1}.
\end{equation}
Only the piece
\begin{equation}
u\left[\kappabf\cdot(\bv_{\nabla B,i} + \bv_{E,0})\right]^{(0)}\partial_u \check F_{i1} \sim \rho_{i*}\frac{v_{ti}}{L_0}  \delta \check F_{i},
\end{equation}
corresponding to the omnigeneous magnetic field $\bB_0$, enters
\eq{eq:DKEtrappedOepsilong}. The effect of higher-order terms like
\begin{equation}\label{eq:termbecomeslarge}
u\left[\kappabf\cdot(\bv_{\nabla B,i} + \bv_{E,0})\right]^{(1)}\partial_u \check F_{i1}
\end{equation}
has not been included. In Section
\ref{sec:solutionDKEverylowcollisionality} we learnt that transport is
dominated by two collisional layers when $\nu_{i*}\ll \rho_{i*}$. In
these layers, derivatives with respect to $u$ (or, equivalently, with
respect to $\lambda$) are large, and they grow as $\nu_{i*}$
decreases. Let us denote by $\Delta_u$ the width of the layer in the
coordinate $u$. The term \eq{eq:termbecomeslarge} becomes comparable
with the pitch-angle scattering piece of the collision operator when
\begin{equation}
  \frac{\delta \rho_{i*} v_{ti}}{L_0(\Delta_u/v_{ti})} \sim \frac{\nu_{ii}}
{({\Delta_u/v_{ti}})^2}.
\end{equation}
If the stellarator is in the $\sqrt{\nu}$ regime, the boundary layer
has a width $\Delta_u^{\sqrt{\nu}}/v_{ti} \sim \sqrt{\nu_{i*}/\rho_{i*}}$ and we
get the estimation~\cite{Shaing2009a}
\begin{equation}
\nu_{\delta *} \sim \delta^2\rho_{i*}.
\end{equation}
If the stellarator is in the superbanana-plateau regime, the size of
the boundary layer is $\Delta^{{\rm sb-p}}_u/v_{ti} \sim
(\nu_{i*}/\rho_{i*})^{1/3}$ and we get~\cite{Shaing2009b}
\begin{equation}\label{eq:nudelta_sb-p}
\nu_{\delta *} \sim \delta^{3/2}\rho_{i*}.
\end{equation}

When $\nu_{i*}\lesssim \nu_{\delta *}$, effects like those described in
\cite{Mynick83} must be taken into account. We leave this for future
work.

\section{Conclusions}
\label{sec:conclusions}

Omnigeneity is the property of stellarators that have been perfectly
optimized regarding neoclassical transport. It has been argued in
\cite{Parra2014} and in the Introduction of the present paper that, in practice, deviations from omnigeneity have a non-negligible effect
on the neoclassical fluxes. It is natural to expect that this effect
will be larger at low collisionality $\nu_{i*}$.

The $1/\nu$ regime in stellarators close to omnigeneity is
studied in \cite{Parra2014}; this regime is defined by
$\rho_{i*}\ll \nu_{i*} \ll 1$. In the core of hot stellarator plasmas, even
lower collisionality regimes are relevant. The subject of this paper
has been the study of the parameter range $\nu_{\delta*} \lesssim \nu_{i*} \lesssim \rho_{i*}$, with
the restriction \eq{eq:smallderivativeslandalpha} for the
perturbations of the omnigeneous configuration (i.e. the gradients of
the perturbations have to be small).

When $\nu_{i*} \lesssim \rho_{i*}$, the components of the drifts tangential to the flux surface
have to be retained. For a generic stellarator in this collisionality regime, the drift-kinetic equation becomes radially non-local. Transport in a stellarator close to omnigeneity conserves radial locality. The appropriate radially local drift-kinetic equation to solve
for the dominant non-omnigeneous piece of the distribution function has
been derived in Section \ref{sec:ExpansionDKEclosetoOmnigeneity}. In
Section \ref{sec:solutionDKEverylowcollisionality} the equation has
been solved and an explicit formula for the ion energy flux $Q_i$ has
been provided in \eq{eq:Qiadditiveformula}. The formula manifests, in
particular, that when $\nu_{i*} \ll \rho_{i*}$ transport is determined
by two small collisional layers. One of the layers is
located around the boundary between trapped and passing particles and
the other is located in the neighborhood of the phase-space points where the
precession frequency (which is caused by the tangential drifts) vanishes. The former corresponds to the
$\sqrt{\nu}$ regime and the latter to the
superbanana-plateau regime. In addition, we have shown that the neoclassical fluxes scale with the square of the size of the deviation from omnigeneity.

In Section \ref{sec:solutionDKEverylowcollisionality} we have also discussed the quasineutrality equation, employed to find the electric field tangent to the flux surface. We have proven that the superbanana-plateau layer needs to be worked out in order to calculate the tangent electric field. The careful analysis of the quasineutrality equation showed that the specific form of the drift-kinetic equation in the superbanana-plateau regime depends on the size of the radial electric field. In Section \ref{sec:varphi1andvarphi0} we have given the equation to determine the radial electric field.

Finally, in Section \ref{sec:Estimationnucritical} we have explained
why the results of Section \ref{sec:solutionDKEverylowcollisionality}
are not valid below a certain value of the collisionality, that we
call $\nu_{\delta *}$ and that we have estimated. The treatment of
the regime $\nu_{i*}\lesssim \nu_{\delta *}$ in stellarators close
to omnigeneity is left for future work.

\appendix

\section{Proof of  relations \eq{eq:DriftpsiIntermsofJ} 
and \eq{eq:DriftalphaIntermsofJ}}
\label{sec:relationJandDrifts}

Starting from \eq{eq:magneticdrift} and \eq{eq:ExBdrift0}, let us first manipulate the radial components of the drifts. We employ that the radial $\nabla B$ drift can be conveniently rewritten with the help of the identity
\begin{equation}\label{eq:rewriteradialnablaBdrift}
(\bun\times\nabla B)\cdot\nabla\psi = \frac{B}{\Psi'_t} \left(
-\partial_\alpha B + \partial_\alpha\boldr\cdot\bun\,\partial_l B\right),
\end{equation}
where we have used $\bB = \Psi'_t\nabla\psi\times\nabla\alpha$, $\partial_l \boldr = \bun$, and the relations
\begin{eqnarray}
\nabla \psi \times\nabla\alpha = \frac{1}{\sqrt{g}}\partial_l\boldr \mbox{ \ \ (and cyclic permutations of $\{\psi,\alpha,l\}$}),
\end{eqnarray}
with the volume element given by
\begin{equation}
\sqrt{g} = \frac{\Psi'_t}{B}.
\end{equation}
Here, the position in euclidean coordinates is viewed as a function of the flux coordinates, $\boldr(\psi,\alpha,l)$.

In order to recast the radial curvature drift we use that $\kappabf = \bun\cdot\nabla\bun$ and that, trivially,
\begin{eqnarray}
\fl
\bun\cdot\nabla\bun = (\bun\cdot\nabla\bun\cdot\partial_\psi\boldr)\nabla\psi + (\bun\cdot\nabla\bun\cdot\partial_\alpha\boldr)\nabla\alpha + (\bun\cdot\nabla\bun\cdot\partial_l\boldr)\nabla l.
\end{eqnarray}
The last term in the previous equation equals zero because $\partial_l \boldr = \bun$ and $\bun\cdot\nabla\bun\cdot\bun \equiv 0$. Then, it is easy to see that
\begin{eqnarray}
(\bun\times\kappabf)\cdot\nabla\psi =
-\frac{B}{\Psi'_t} \bun\cdot\nabla\bun\cdot\partial_\alpha\boldr.
\end{eqnarray}
Noting that $\bun\cdot\nabla\bun = \partial_l \bun $, integrating by parts in $l$ and using that $\bun\cdot\partial_\alpha\partial_l\boldr = \bun\cdot\partial_\alpha\bun\equiv 0$, we get
\begin{eqnarray}\label{eq:rewriteradialcurvaturedrift}
(\bun\times\kappabf)\cdot\nabla\psi =
-\frac{B}{\Psi'_t} \partial_l (\bun\cdot\partial_\alpha\boldr).
\end{eqnarray}

Finally, the radial $E\times B$ drift will be rewritten by employing
\begin{equation}\label{eq:rewriteradialExBdrift}
(\bun\times\nabla\varphi)\cdot\nabla\psi = 
\frac{B}{\Psi'_t} \left(
-\partial_\alpha \varphi + \partial_\alpha\boldr\cdot\bun\,\partial_l \varphi\right),
\end{equation}
which is obtained exactly in the same way as \eq{eq:rewriteradialnablaBdrift}.

Recalling \eq{eq:magneticdrift} and \eq{eq:ExBdrift0}, and collecting the results \eq{eq:rewriteradialnablaBdrift}, \eq{eq:rewriteradialcurvaturedrift} and \eq{eq:rewriteradialExBdrift}, we find
\begin{eqnarray}\label{eq:DriftpsiIntermsofJAux}
\fl  2\int_{l_{b_1}}^{l_{b_2}}\frac{1}{|v_{||}|}(\bv_{M,i}+\bv_{E})
\cdot\nabla\psi \, \dd l=
\nonumber\\[5pt]
\fl\hspace{1cm}
\frac{2m_ic}{Z_ie\Psi'_t}\partial_\alpha \int_{l_{b_1}}^{l_{b_2}}|v_{||}|\dd l
-
\frac{2m_ic}{Z_ie\Psi'_t}
\int_{l_{b_1}}^{l_{b_2}}\partial_l(|v_{||}|\partial_\alpha\boldr\cdot\bun)\dd l.
\end{eqnarray}

Analogously, one can show that
\begin{eqnarray}\label{eq:DriftalphaIntermsofJAux}
\fl  2\int_{l_{b_1}}^{l_{b_2}}\frac{1}{|v_{||}|}(\bv_{M,i}+\bv_{E})
\cdot\nabla\alpha \, \dd l=
\nonumber\\[5pt]
\fl\hspace{1cm}
-\frac{2m_ic}{Z_ie\Psi'_t}\partial_\psi \int_{l_{b_1}}^{l_{b_2}}|v_{||}|\dd l
+
\frac{2m_ic}{Z_ie\Psi'_t}
\int_{l_{b_1}}^{l_{b_2}}\partial_l(|v_{||}|\partial_\psi\boldr\cdot\bun)\dd l.
\end{eqnarray}
The last term in both \eq{eq:DriftpsiIntermsofJAux} and
\eq{eq:DriftalphaIntermsofJAux} vanishes because $v_{||}$ equals zero
at $l_{b_1}$ and $l_{b_2}$. Finally, using definition
\eq{eq:defJ}, we obtain \eq{eq:DriftpsiIntermsofJ} and
\eq{eq:DriftalphaIntermsofJ}.

\section{Proof that the adiabatic response does not contribute to the energy flux}
\label{sec:adiabaticresponsedoesnotcontribute}
The adiabatic response is contained in $F_{i0}^{\cE (0)}$, defined in \eq{eq:Fi0^(0)E}. Its contribution to the energy flux \eq{eq:defenergyfluxIntro} is given by
\begin{eqnarray}\label{eq:Qiadiab}
\fl Q_{i, {\rm ad}} = 4\pi V'(\psi)
\left\langle
\int_{Z_i e\varphi / m_i}^\infty
\dd\cE\int_0^{B^{-1}(\cE - Z_i e\varphi / m_i)}
\dd\mu \, \frac{B}{|v_{||}|}m_i \cE \, \bv_{d,i}\cdot\nabla\psi F_{i0}^{\cE (0)}
\right\rangle_\psi,
\end{eqnarray}
with $\bv_{d,i} = \bv_{M,i} + \bv_E$ (see definitions \eq{eq:magneticdrift} and \eq{eq:ExBdrift0}). In \eq{eq:Qiadiab} we have used that
\begin{equation}
\int f \, \dd^2 S = V'(\psi)\langle f |\nabla\psi| \rangle_\psi,
\end{equation}
where the flux surface average operation and $V'(\psi)$ have been defined in \eq{eq:def_fluxsurfaceaverage} and \eq{eq:def_volume}.

A direct check shows that
\begin{equation}
\bv_{d,i}\cdot\nabla\psi = \frac{v_{||}}{\Omega_i}\nabla\cdot(v_{||}\bun\times\nabla\psi).
\end{equation}
Then,
\begin{eqnarray}\label{eq:Qiadiab2}
\fl Q_{i, {\rm ad}} =  \frac{4\pi m_i^2 c}{Z_i e}V'(\psi)
\left\langle
\int_{Z_i e\varphi / m_i}^\infty
\dd\cE\int_0^{B^{-1}(\cE - Z_i e\varphi / m_i)}
\dd\mu \, 
\nabla\cdot \left(
\cE
F_{i0}^{\cE (0)}
|v_{||}|\bun\times\nabla\psi
\right)
\right\rangle_\psi
\nonumber\\[5pt]
\fl
\hspace{0.5cm}
=
\frac{4\pi m_i^2 c}{Z_i e}V'(\psi)
\left\langle
\nabla\cdot
\int_{Z_i e\varphi / m_i}^\infty
\dd\cE\int_0^{B^{-1}(\cE - Z_i e\varphi / m_i)}
\dd\mu \, 
 \left(
\cE
F_{i0}^{\cE (0)}
|v_{||}|\bun\times\nabla\psi
\right)
\right\rangle_\psi
\end{eqnarray}
where we have used that $F_{i0}^{\cE (0)}$ is a flux function, and in the second equality we have employed that the integrand vanishes when $\mu = B^{-1}(\cE - Z_i e\varphi / m_i)$ and when $\cE = Z_i e\varphi / m_i$. Finally, recalling the identity
\begin{equation}
\left\langle
\nabla\cdot\mathbf{A}
\right\rangle
= \frac{1}{V'(\psi)}\partial_\psi
\left(
V' (\psi)\left\langle
\mathbf{A}\cdot\nabla\psi
\right\rangle_\psi
\right)
\end{equation}
for any vector field $\mathbf{A}$ and applying it to \eq{eq:Qiadiab2} we deduce that $Q_{i, {\rm ad}}$ vanishes.

\section{Asymptotic expansion of $\partial_\psi J^{(0)}$ near the
  boundary between trapped and passing particles}
\label{sec:Expansionj0psi}

We show that
\begin{equation}
  \partial_\psi J^{(0)}
  =
  -\int_{l_{b_{10}}}^{l_{b_{20}}}
\frac{\lambda v\partial_\psi B_0 + 2Z_ie/(m_iv)\partial_\psi\varphi_0}
{\sqrt{1-\lambda B_0}}\dd l
\end{equation}
has the form \eq{eq:expansionj0psi} for
small $\lambda-\lambda_c > 0$ by, first, using the trivial identity
\begin{eqnarray}\label{eq:rewritetaub}
\fl
\partial_\psi J^{(0)} = 
- \sum_{k=1}^2\int_{l_{b_{10}}}^{l_{b_{20}}}
\frac{\lambda_c v\partial_\psi B_0(l_{M,k}) + 2Z_ie/(m_iv)\partial_\psi\varphi_0}
{\sqrt{(\lambda_c/2)|\partial_l^2B_0(l_{M,k})|(l-l_{M,k})^2
- B_0(l_{M,k}) (\lambda - \lambda_c)}}\dd l
\nonumber\\[5pt]
\fl\hspace{1cm}
-
 \int_{l_{b_{10}}}^{l_{b_{20}}}
\Bigg(\frac{\lambda v\partial_\psi B_0(l)+ 2Z_ie/(m_iv)\partial_\psi\varphi_0}{\sqrt{1-\lambda B_0(l)}}
\nonumber\\[5pt]
\fl\hspace{1cm}
-\sum_{k=1}^2
\frac{\lambda_c v\partial_\psi B_0(l_{M,k})+ 2Z_ie/(m_iv)\partial_\psi\varphi_0}
{\sqrt{(\lambda_c/2)|\partial_l^2B_0(l_{M,k})|(l-l_{M,k})^2
- B_0(l_{M,k}) (\lambda - \lambda_c)}}
\Bigg)\dd l,
\end{eqnarray}
which is well defined for sufficiently small
$\lambda-\lambda_c$. Here, we have only displayed the
dependence of $B_0$ on $l$. The values $l_{M,k}$, for $k=1,2$, locate
two consecutive absolute maxima of $B_0$ when moving along the field
line; in particular, $B_0(l_{M,k}) = B_{0,\max}$ for $k=1,2$. The
second integral on the right side of \eq{eq:rewritetaub} is finite
when $\lambda \to \lambda_c$, and hence it contributes to $a_2$ and
higher-order terms in \eq{eq:expansionj0psi}. The first integral on
the right side of \eq{eq:rewritetaub} can be computed analytically;
namely,
\begin{eqnarray}\label{eq:analyticalexpression}
\fl  - \sum_{k=1}^2 
\int_{l_{b_{10}}}^{l_{b_{20}}}
  \frac{\lambda_c v\partial_\psi B_0(l_{M,k})+ 2Z_ie/(m_iv)\partial_\psi\varphi_0}
{\sqrt{(\lambda_c/2)|\partial_l^2B_0(l_{M,k})|(l-l_{M,k})^2
      - B_0(l_{M,k}) (\lambda - \lambda_c)}}\dd l
  =
\nonumber\\[5pt]
\fl\hspace{0.5cm}
  -
\sum_{k=1}^2
\frac{\lambda_c v\partial_\psi B_0(l_{M,k})+ 2Z_ie/(m_iv)\partial_\psi\varphi_0}
{\sqrt{(\lambda_c/2)|\partial_l^2B_0(l_{M,k})|}}
\nonumber\\[5pt]
\fl\hspace{0.5cm}
\times\Bigg[
  \ln\Bigg(|x|+\sqrt{x^2-\frac{2B_0(l_{M,k})}
{\lambda_c |\partial_l^2B_0(l_{M,k})|}(\lambda-\lambda_c)}\,\Bigg)
\Bigg]_{x\, = \, l_{b_{10}} - l_{M,k}}^{x\, = \, l_{b_{20}} - l_{M,k}}.
\end{eqnarray}

For small $\lambda - \lambda_c$,
\begin{equation}
l_{b_{10}} - l_{M,1} 
= \sqrt{\frac{2B_0(l_{M,k})(\lambda-\lambda_c)}
{\lambda_c|\partial_l^2B_0(l_{M,k})|}} + \dots,
\end{equation}
and
\begin{equation}
l_{b_{20}} - l_{M,2} 
= - \sqrt{\frac{2B_0(l_{M,k})(\lambda-\lambda_c)}
{\lambda_c|\partial_l^2B_0(l_{M,k})|}} + \dots,
\end{equation}
whereas $l_{b_{20}} - l_{M,1} = O(L_0)$ and $l_{b_{10}} - l_{M,2} =
O(L_0)$. Using these results in \eq{eq:analyticalexpression}, it is
straightforward to deduce that
\begin{eqnarray}
\fl  - \sum_{k=1}^2
\int_{l_{b_{10}}}^{l_{b_{20}}}
  \frac{\lambda_c v\partial_\psi B_0(l_{M,k})+ 2Z_ie/(m_iv)\partial_\psi\varphi_0}{\sqrt{(\lambda_c/2)|
\partial_l^2B_0(l_{M,k})|(l-l_{M,k})^2
      - B_0(l_{M,k}) (\lambda - \lambda_c)}}\dd l
  =
\nonumber\\[5pt]
\fl\hspace{0.5cm}
 \sqrt{\frac{1}{2\lambda_c}}
\sum_{k=1}^2
  \frac{\lambda_c v\partial_\psi B_0(l_{M,k})+ 2Z_ie/(m_iv)\partial_\psi\varphi_0}{\sqrt{|
\partial_l^2B_0(l_{M,k})|}}
  \ln(B_{0,\max}(\lambda-\lambda_c))\, + O(v_{ti} L_0 / \psi),
\end{eqnarray}
from which equation \eq{eq:a1} follows.

\section{Analysis of equation
  \eq{eq:boundarylayerInfFourier} in a neighborhood of $\lambda = \lambda_c$}
\label{sec:IrregularSingularPoints}

 In this appendix we use the variable $x = \lambda -
\lambda_c$ and rewrite \eq{eq:boundarylayerInfFourier} as
\begin{eqnarray}\label{eq:boundarylayerInfFourierApp}
\partial^2_x
g_n
+ i n \frac{a_1}{\nu_\lambda\xi} \ln ({\tilde a}_2 x) \, g_n
=
-\partial^2_x \underline{g_{0,n}},
\end{eqnarray}
where $g_n(x) = { g}_{{\rm bl},n}(\lambda_c + x)$ and
$\underline{g_{0,n}}(x) = { g}_{0,n}(\lambda_c + x)$. The equations
\eq{eq:boundarylayerInfFourier} for $n\neq 0$ (recall that $g_0 (x)$
and $\underline{g_{0,0}}(x)$ vanish) have an irregular singular
point~\cite{BenderOrszag} at $x = 0$.

The standard methods do not
work when applied to the
homogeneous equation
\begin{eqnarray}\label{eq:boundarylayerInfFourierAppHom}
\partial^2_x
g_n
+ i n \frac{a_1}{\nu_\lambda\xi} \ln ({\tilde a}_2 x) \, g_n
=
0
\end{eqnarray}
near $x=0$. However, one can check that the ansatz
\begin{equation}
g_n = \sum_{m,p = 0}^\infty
A_{m,p}\, x^{2p + m}(\ln x)^p
\end{equation}
is consistent, in the sense that by substitution in
\eq{eq:boundarylayerInfFourierAppHom} one can find recurrence
relations that determine all the coefficients $A_{m,p}$ except two of
them. The free coefficients can be taken to be $A_{0,0}$ and
$A_{1,0}$. In order to show this, it is advisable to start by writing the equation provided by terms in \eq{eq:boundarylayerInfFourierAppHom} that are proportional to $\ln x$ and the equation corresponding to terms proportional to $x\ln x$. 

Hence, there exist two linearly
independent solutions of \eq{eq:boundarylayerInfFourierAppHom} that
are finite at $x=0$.

It is easy to realize that the source term on the right side of
\eq{eq:boundarylayerInfFourierApp} does not make $g_n$ diverge at
$x=0$. First, note that $\underline{g_{0,n}}$ is finite for any value of $x$. If
one takes $g_n = -\underline{g_{0,n}} + f_n$, \eq{eq:boundarylayerInfFourierApp}
gives the following equation for $f_n$:
\begin{eqnarray}\label{eq:boundarylayerInfFourierAppAux}
\partial^2_x
f_n
+ i n \frac{a_1}{\nu_\lambda\xi} \ln ({\tilde a}_2 x) \, f_n
=  \frac{i n}{\nu_\lambda\xi}
\left(
c_{1,n}\ln ({\tilde a}_2 x) + {\tilde c}_{2,n}
\right)
\Upsilon_i F_{i0}
,
\end{eqnarray}
where $c_1$ and ${\tilde c}_2$ have been defined in
\eq{eq:c1} and \eq{eq:c2tilde}. Since the indefinite
integrals of $\ln x$ are finite everywhere, the source term on the
right side of \eq{eq:boundarylayerInfFourierAppAux} does not introduce
singularities in $f_n$ and we conclude that $g_n$ is finite for any
value of $x$; in particular, it is finite at $x=0$.

\ack

This work has been carried out within the framework of the EUROfusion
Consortium and has received funding from the Euratom research and
training programme 2014-2018 under grant agreement No 633053. The
views and opinions expressed herein do not necessarily reflect those
of the European Commission. This research was supported in part by grants ENE2012-30832 and ENE2015-70142-P, Ministerio de Econom\'{\i}a y Competitividad,
Spain.


\end{document}